\RequirePackage{rotating} 
\documentclass[format=acmsmall, review=false, screen=true]{acmart}

\usepackage{booktabs} 

\usepackage[ruled]{algorithm2e} 

\SetAlFnt{\small}
\SetAlCapFnt{\small}
\SetAlCapNameFnt{\small}
\SetAlCapHSkip{0pt}
\IncMargin{-\parindent}

\usepackage{verbatim}
\usepackage{multicol}
\usepackage{multirow}
\usepackage{graphicx}
\usepackage{rotating}
\usepackage{makecell}

\usepackage{color}
\usepackage{listings}
\usepackage[tight,footnotesize]{subfigure}

\setcopyright{acmlicensed}

\acmDOI{0000001.0000001}

\begin{document}

\title[ContextServ: Towards Model-Driven Development of Context-Aware Web Services]{ContextServ: Towards Model-Driven Development of Context-Aware Web Services}

\author{QUAN Z. SHENG}
\orcid{0000-0002-3326-4147}
\affiliation{%
  \institution{Macquarie University}
  \streetaddress{104 Jamestown Rd}
  \city{Sydney}
  \state{NSW}
  \postcode{2109}
  \country{Australia}}

\author{JIAN YU}
\affiliation{%
  \institution{Auckland University of Technology}
  \streetaddress{55 Wellesley St E}
  \city{Auckland}
  \postcode{1010}
  \country{New Zealand}}
\author{HANCHUAN XU}
\affiliation{%
  \institution{Harbin Institute of Technology}
  \city{Haerbin}
  \postcode{150001}
  \country{China}}  
 \author{WEI EMMA ZHANG}
\orcid{0000-0002-0406-5974}
\affiliation{%
  \institution{Macquarie University}
  \streetaddress{104 Jamestown Rd}
  \city{Sydney}
  \state{NSW}
  \postcode{2109}
  \country{Australia}} 
 \author{ANNE H.H. NGU}
\affiliation{%
  \institution{Texas State University}
  \streetaddress{601 University Dr}
  \city{San Marcos}
  \state{TX}
  \postcode{78666}
  \country{USA}} 
 \author{JUN HAN}
\affiliation{%
  \institution{Swinburne University of Technology}
  \streetaddress{John St}
  \city{Melbourne}
  \state{VIC}
  \postcode{3122}
  \country{Australia}} 
 \author{RUILIN LIU}
\affiliation{%
  \institution{Harbin Institute of Technology}
  \city{Haerbin}
  \postcode{150001}
  \country{China}} 

\begin{abstract}
In the era of Web of Things and Services, Context-aware Web Services (CASs) are emerging as an important technology for building innovative context-aware applications. 
CASs enable the information integration from both the physical and virtual world, which affects human living.   
However, it is challenging to build CASs, due to the lack of context provisioning management approach and limited generic approach for formalizing the development process. 
We therefore propose ContextServ, a platform that uses a model-driven approach to support the full life cycle of CASs development, hence offering significant design and management flexibility. 
ContextServ implements a proposed UML-based modelling language ContextUML to support multiple modelling languages. It also supports dynamic adaptation of WS-BPEL based context-aware composite services by weaving context-aware rules into the process. 
Extensive experimental evaluations on ContextServ and its components showcase that ContextServ can support effective development and efficient execution of context-aware Web services. 

\end{abstract}

%
%

%
%


\thanks{This work is supported by the Australian Research Council, under grant DP0878367. 
   
  Authors' address:  Q. Z. Sheng, Department of Computing, Macquarie University, Sydney, NSW 2109, Australia;
  J. Yu, Department of Computer Science, Auckland University of Technology, 55 Wellesley St E, Auckland 1010, New Zealand;
  H. Xu and R. Liu, School of Computer Science and Technology, Harbin Institute of Technology, Haerbin 150001, China; 
  W. E. Zhang, Department of Computing, Macquarie University Sydney, NSW 2109, Australia;
  A. H.H. Ngu, Department of Computer Science, Texas State University, 601 University Dr, San Marcos, TX 78666, USA;
  J. Han, Faculty of Information and Communication Technologies, Swinburne University of Technology, John St, Melbourne, VIC 3122, Australia
}

\maketitle

\renewcommand{\shortauthors}{Q. Z. Sheng et al.}

\section{Introduction}
Over the years, the Web has gone through many transformations, from traditional linking and sharing of computers and documents (i.e. ``Web of Data") to current connecting of people (i.e. ``Web of People") . With the recent advances in radio-frequency identification technology, sensor networks, and web services, the web is continuing the transformation and will be slowly evolving into the so-called ``Web of Things and Services" \cite{Sheng-book2010,Stankovic14,RazzaqueMPC16}. Indeed, this future Web will provide an environment where everyday physical objects such as buildings, sidewalks, and commodities are readable, recognizable, addressable, and even controllable using services via the web. The capability of integrating the information from both the physical world and the virtual one not only affects the way how we live, but also creates tremendous new Web-based business opportunities such as support of independent living of elderly persons, intelligent traffic management, efficient supply chains, and improved environmental monitoring. Therefore, context awareness, which refers to the capability of an application or a service being aware of its physical environment or situation (i.e., context) and responding proactively and intelligently based on such awareness \cite{Abowd-02,Dey-TOCHI05,Julien-TSE06}, has been identified as one of the key challenges and most important trends in computing today 
and 
holds the potential to make our daily lives more productive, convenient, and enjoyable. 

Nowadays, Web services have become a major technology to implement loosely coupled business processes and perform application integration. Through the use of context, a new generation of smart Web services is currently emerging as an important technology for building innovative context-aware applications. We call such category of Web services as context-aware web services (CASs). A CAS is a Web service that uses context information to provide relevant information and/or services to users \cite{Julien-TSE06,Kapitsakietal.2009,Sheng-book2010,PereraLJC15}. A CAS can present relevant information or can be executed or adapted automatically, based on available context information. For instance, a tour-guide service gives tourists suggestions on the attractions to visit by considering their current locations, preferences, and even the prevailing weather conditions \cite{Liao-iiWAS2009}. CASs are emerging as an important technology to underpin the development of new applications (user centric, highly personalized) on the future ubiquitous web.

Although the combination of context awareness and Web services sounds appealing, injecting context into Web services raises a number of significant challenges, which have not been widely recognized or addressed by the Web services community \cite{Sheng-book2010,yu2008deploying,ShengQVSBX14}. One reason for this difficulty is that current Web services standards, such as the Web Services Description Language (WSDL), Web Application Description Language (WADL),  and the Simple Object Access Protocol (SOAP), are not sufficient for describing and handling context information. CAS developers must implement everything related to context management, including collection, dissemination, and usage of context information, in an ad hoc manner. Another reason is that CASs are frequently required to be dynamically adaptive in order to cope with constant changes, which means a service being able to change its behaviour at runtime in accordance with the contexts. Unfortunately, service-oriented systems built with WS-BPEL (Web Services Business Process Execution Language) are still too rigid. The third reason is, to the best of our knowledge, there is a lack of generic approaches for formalizing the development of CASs. As a consequence, developing and maintaining CASs is a very cumbersome, error-prone, and time consuming activity, especially when these CASs are complex.

Motivated by these concerns, we have developed the ContextServ platform for rapid development of CASs. One innovative feature of ContextServ is to use a model-driven approach that offers significant design flexibility by separating the modelling of context and context awareness from service components, which eases both development and maintenance of CASs. The second feature of ContextServ is that it supplies a set of automated tools for generating and deploying executable implementations of CASs. ContextServ supports the full life cycle of 
CASs 
development, 
including a visual ContextUML editor, a ContextUML to WS-BPEL translator, and a WS-BPEL deployer. 
The third feature of ContextServ is that it supports dynamic adaptation of WS-BPEL based context-aware composite services by weaving context-aware rules into the process. 

This paper is a comprehensive summary of our previous \cite{Sheng-ICMB05,Sheng-ICSE09,sheng2010techniques,yu2015model} and latest research results. The major contributions of this paper are:
\begin{itemize}
	\item The introduction of our proposed ContextUML, an UML-based modelling language, and the CotextServ platform we developed to implement ContextUML for model-driven development of CASs,
	\item Optimization of Context Service Community which is for dynamic and optimized context provisioning for CASs, 
	\item A detailed report of the MoDAR (\underline{Mo}del-Driven Development of \underline{D}ynamically Adaptive Service-Oriented Systems with \underline{A}spects and \underline{R}ules) approach which provides support to development of dynamically adaptive WS-BPEL based CASs,
	\item An extensive evaluation of the ContextServ, including a controlled usability test for ContextServ and two performance studies on the Context Service Community and the MoDAR approach. 
\end{itemize}

The rest of this paper is organized as follows. Section \ref{sec:ContextUML} introduces the ContextUML language and Section \ref{sec:contextserv} introduces the architecture and implementation of the ContextServ platform. The MoDAR approach 
is introduced in Section \ref{sec:MoDAR}. Section \ref{sec:community1} presents the design of context service community, an important concept for optimized selection of context information for CASs. Section \ref{sec:Eval} evaluates the usability of ContextServ development environment and performances of context service community and MoDAR. Section \ref{sec:relatedwork} surveys the related work and compares them with our approach. Finally, Section \ref{sec:conclusion} concludes the paper.

\section{ContextUML}
\label{sec:ContextUML}
\begin{figure}[]
	\centering
	\includegraphics[width=\linewidth]{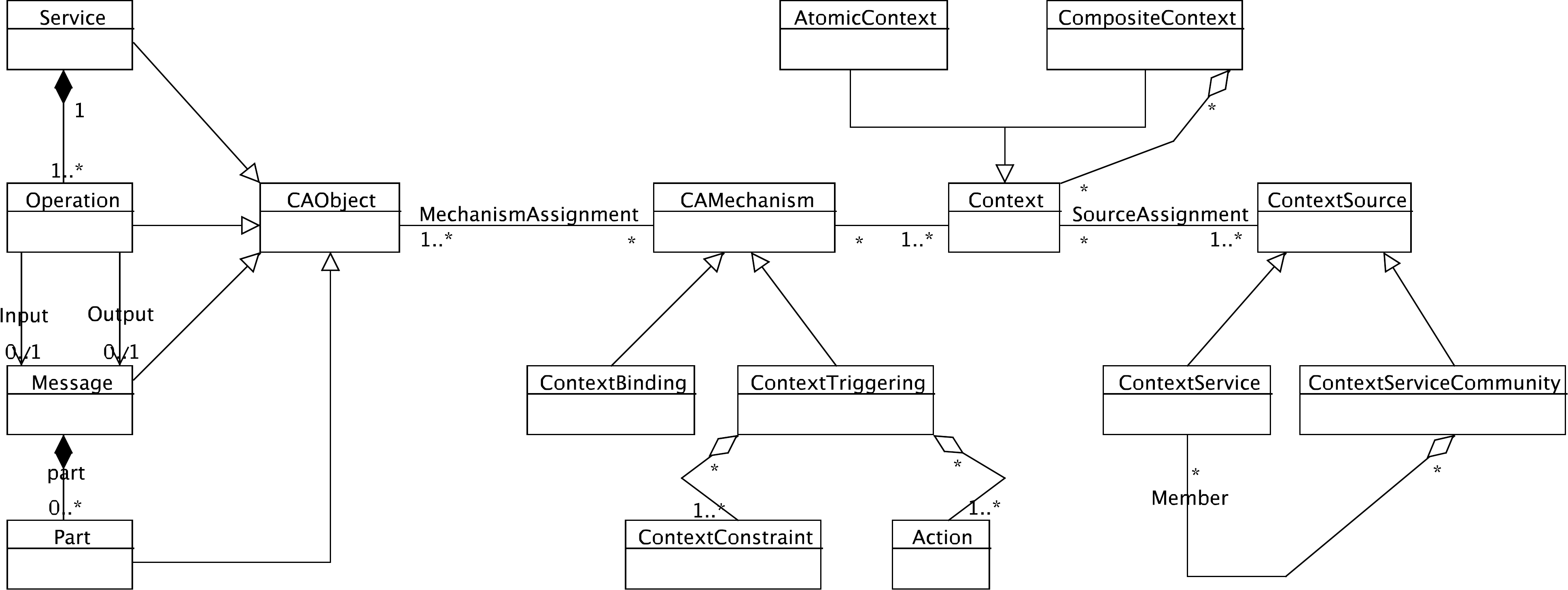}
	\caption{ContextUML metamodel}
	\label{fig:ContextUML}
\end{figure}

In this section, we first introduce ContextUML, a Unified Modeling Language (UML) based language for model-driven development of context-aware Web services~\cite{Sheng-ICMB05}. ContextUML metamodel is shown in Figure~\ref{fig:ContextUML}, which can be divided into two parts: {\em context modelling metamodel} and {\em context-awareness modelling metamodel}.

\subsection{Context Modelling}
\subsubsection{Context Type}
\label{sec:contexttype}

{\small \sf Context} is a class that models the context information. 
In our design, the type {\small \sf Context} is further 
distinguished into
 two categories that are formalized by the subtypes 
 {\small \sf AtomicContext} and {\small \sf CompositeContext}. 
Atomic contexts
 are low-level contexts that do not rely on other contexts 
 and can be provided directly 
 by context sources.
In contrast, composite contexts are high-level contexts that may not have 
direct counterparts on the context provision. A composite context 
aggregates multiple contexts, either atomic or composite.
The concept of composite context  
can be used to 
provide a rich modelling vocabulary. 

For instance, in the scenario of tour-guide, 
{\tt temperature} and {\tt wind speed} are atomic contexts because they 
can be provided by a local weather forecast Web service.  
Whereas, {\tt harshWeather} is a composite context that aggregates the former two contexts.

\subsubsection{Context Source}
\label{sec:source}
The 
{\small \sf ContextSource} models the resources from which contexts are retrieved.  
We abstract two categories of context sources, formalized by the context source subtypes 
{\small \sf ContextService} and {\small \sf ContextServiceCommunity}, respectively. 
A context service is provided by an autonomous organization (i.e.,
context provider), collecting, refining, and disseminating context information. 
To solve the challenges of heterogeneous and dynamic context information, we 
abstract the concept of context service community, 
which enables the dynamic provisioning of
optimal contexts. 
The concept is evolved from {\em service community} we developed in~\cite{SELFSERV-DAPD} 
and details will be given in Section~\ref{sec:community}. 

It should be noted that in ContextUML, we do not model the acquisition 
of context information, such as how to collect {\em raw} context
information from sensors. Instead, context services that we abstract in
ContextUML encapsulate sensor details and provide context information by 
interpreting and transforming the sensed information (i.e., raw context 
information). The concept of context service 
hides the complexity of context acquisition from CAS
designers so that they can focus on the functionalities of CASs, rather than context
sensing. 

\subsubsection{Context Service Community}
\label{sec:community}

A context service community aggregates multiple context services, offering with a 
unified interface. It is intended as a means to support the dynamic retrieval of 
context information. A community describes the capabilities of a desired service 
(e.g., providing user's location) without referring to any actual context service 
(e.g., {\tt WhereAmI} service). When the operation of a community is invoked, 
the community is responsible for selecting the most appropriate context service that 
will provide the requested context information. 
Context services can join and leave communities at any time. 

By abstracting
{\small \sf ContextServiceCommunity} as one of 
the 
 context sources, we can enable the dynamic 
context provisioning. In other words, CAS designers do not have
to specify which context services are needed for context information retrieval at 
the design stage. The decision of which specific context 
service should be selected for the 
provisioning of a context is postponed until the invocation of 
CASs. The selection can be based on a 
{\em multi-criteria utility function}~\cite{Stolze2001,SELFSERV-DAPD} and  
the criteria used in the function can be a set of 
{\em Quality of Context} (QoC) parameters~\cite{QoC-03}.

The quality of context is extremely important for CASs in the sense that context information 
is used to automatically adapt services or content they provide. The imperfection of 
context information may make CASs {\em misguide} their users. 
For example, if the weather information is outdated, our attractions searching service 
might suggest users to surf at the Bondi Beach although it is rainy and stormy. Via context 
service communities, the optimal context information is always selected, which in turn, 
ensures the quality of CASs. Since context service community is an important concept for CASs, we have a separate section that gives more details on our design (see Section~\ref{sec:community1}). 

\subsection{Context Awareness Modelling}
{\small \sf CAMechanism} is a class that formalizes the mechanisms for 
context awareness (CA for short). We 
differentiate between two categories of context awareness mechanisms 
by subtypes {\small \sf ContextBinding} 
and {\small \sf ContextTriggering}, which will be detailed in 
Section~\ref{sec:binding} and Section~\ref{sec:trigger}, respectively.   
Context awareness mechanisms are assigned to 
context-aware objects---modelled in the 
type {\small \sf CAObject}---by the relation {\small \sf MechanismAssignment}, 
indicating which objects have what kinds of context awareness mechanisms. 

{\small \sf CAObject} is a base class of all model elements in ContextUML that represent
context-aware objects. There are four subtypes of {\small \sf CAObject}: 
{\small \sf Service}, {\small \sf Operation},  
{\small \sf Message}, and {\small \sf Part}. 
Each service offers one or more operations and 
each operation belongs to exactly one service. The relation 
is denoted by a composite aggregation (i.e., the association end with a filled diamond 
in Figure \ref{fig:ContextUML}). 
Each operation may have one input and/or one output messages. Similarly, 
each message may have multiple parts (i.e., parameters).  
A context awareness mechanism can be assigned to either a service, an operation of a service, 
input/output messages of an operation, or even a particular part (i.e., parameter) 
of a message. 
It is worth mentioning that the four primitives are directly 
adopted from WSDL,
which enables designers to build CASs on top of the previous
implementation of Web services. 

\subsubsection{Context Binding}
\label{sec:binding}
{\small \sf ContextBinding} is a subtype of {\small \sf CAMechanism} 
that models the automatic binding of contexts 
to context-aware objects.
By abstracting the concept of context binding, it is possible to automatically 
retrieve information for users based on available context information. 
For example, suppose that the operation 
of our example CAS
has an input parameter {\tt city}. Everyone who wants to invoke 
the service needs to supply a city name 
to search the attractions.
Further suppose that we have a context  
{\tt userLocation} that represents the city a user is currently in.
A context binding 
can be built between {\tt city} (input parameter of the service) and 
{\tt userLocation} (context). 
The result is that whenever our CAS is invoked, it will  
automatically retrieve attractions in the city where the requester is currently located. 

An automatic contextual reconfiguration (i.e., context binding) 
is actually a {\em mapping} between a context and a context-aware 
object (e.g., an input parameter of a service operation). The semantics is that the 
value of the object is supplied by the value of the context.   
Note that the value of a context-aware object could be 
derived from multiple contexts. For the sake of the simplicity, we restrict our mapping 
cardinality as one to one. In fact, thanks to the introduction of the concept of 
composite context, we can always model an appropriate composite context 
for a context-aware object whose value needs to be derived from multiple contexts.

\subsubsection{Context Triggering}
\label{sec:trigger}
The type {\small \sf ContextTriggering} models the situation of contextual adaptation 
where services can be automatically executed or modified 
based on context information. A context triggering mechanism 
contains two parts: a set of {\em context constraints} and 
a set of {\em actions}, with the semantics of that the actions must be executed 
if and only if all the context constraints are evaluated to true. 

A context constraint specifies that a certain
context must meet certain condition in order to perform a particular operation. 
Formally, a context constraint is modelled as a predicate (i.e., a Boolean function)
that consists of an operator and two or more operands. The first operand always
represents a context, while the other operands may be either constant values
or contexts. An operator can be either a prefix operator that accepts two or
more input parameters or a binary infix operator (e.g., =, $\le$) that 
compares two values. Examples of context constraints can be: 
i) {\tt harshWeather}$=true$; ii) {\texttt windSpeed}$\le 25$. 

Considering our tour-guide application, we can have a context triggering mechanism assigned to its output message. The constraint part of the
mechanism is {\texttt harshWeather}$=true$, and the action part is a transformation function
{\texttt filter}$({\mathcal M}, {\mathcal R})$, where $\mathcal M$ is the output message and 
$\mathcal R$ is a transformation rule (e.g., selecting only indoor attractions).
Consequently, when weather condition is not good, 
the output message will be automatically filtered (e.g., removing outdoor attractions) by the service.

\section{ContextServ Platform}
\label{sec:contextserv}
In this section, we will introduce the ContextServ, a comprehensive platform for simplifying the development of context-aware 
Web services.

\begin{figure}[!t]
	\centering
	\includegraphics[width=11cm]{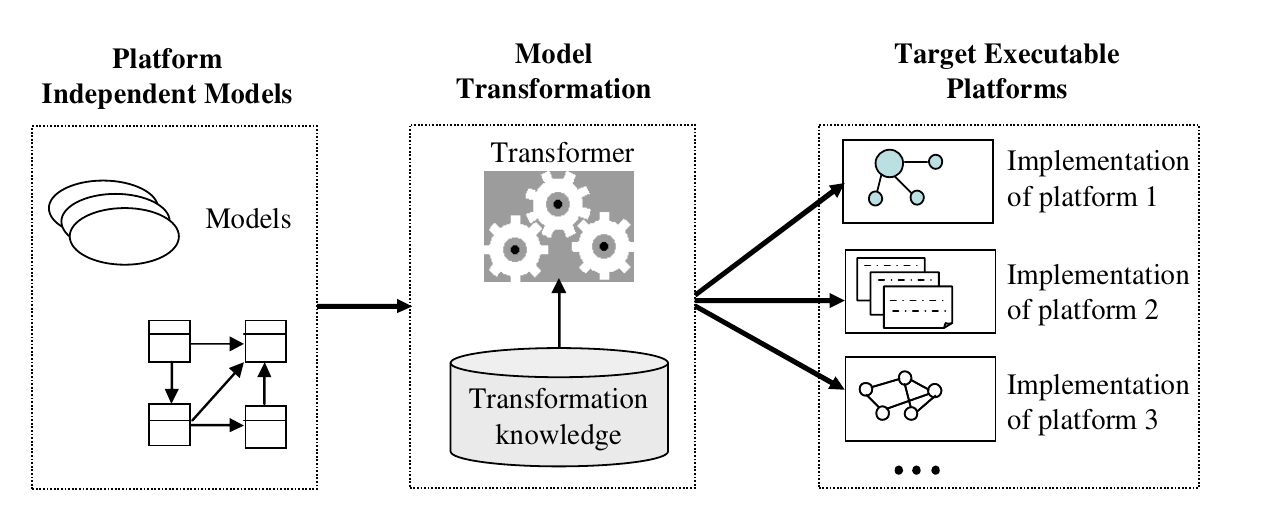}
	\caption{Model-driven development}
	\label{fig:MDD}
\end{figure}

ContextServ adopts model-driven development (MDD)~\cite{frankelBook,mda-03} and the basic idea of MDD is illustrated in Figure~\ref{fig:MDD}. 
Adopting a high-level of abstraction, software systems can be specified in platform independent models (PIMs), which are then (semi-) automatically transformed into platform specific models (PSMs) of target executable platforms using some transformation tools. The same PIM can be transformed into different executable platforms (i.e., multiple PSMs), thus considerably simplifying software development.     

\begin{figure*}[!t]
\centering
\includegraphics[width=0.94\linewidth]{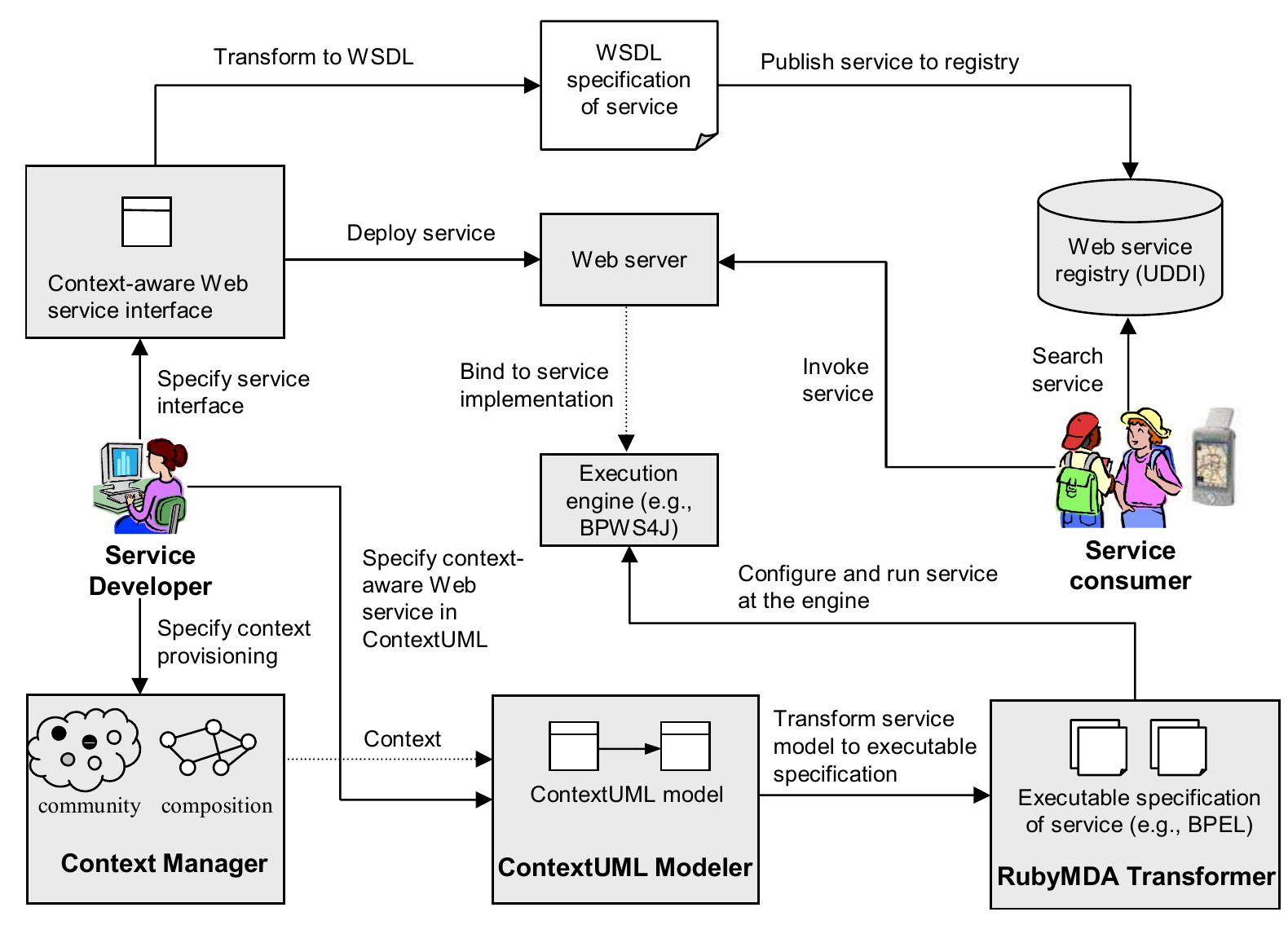}
\caption{Architecture of the ContextServ platform}
\label{fig:architecture}
\end{figure*}

ContextServ relies on ContextUML (see Section~\ref{sec:ContextUML}), a UML-based modeling language that provides high-level, visual 
constructs for specifying context-aware Web services. 
In particular, the language abstracts two context awareness mechanisms,
namely {\em context binding} and 
{\em context triggering}. The former models automatic contextual configuration (e.g., automatic invocation of Web services by mapping a context onto a particular service input parameter), while the latter models contextual adaptation where services can be dynamically modified based on context information. Service models specified in ContextUML are then automatically translated into executable implementations 
(e.g., WS-BPEL specifications) of specific target service implementation platforms (e.g., 
IBM's BPWS4J\footnote{http://www.alphaworks.ibm.com/tech/bpws4j.}). 
Figure \ref{fig:architecture} illustrates the architecture of the ContextServ platform, which consists of three main components, namely {\em Context Manager}, {\em ContextUML Modeller} and {\em RubyMDA Transformer}. We discuss these three components in detail in the following sections.

\subsection{Context Manager}
The context manager provides facilities for service developers to specify context provisioning. Current implementation supports the management of atomic context, composite context, and context community.

\subsubsection{Managing Atomic Contexts and Composite Contexts}
As mentioned before, atomic contexts are low-level contexts that can be obtained directly from context sources. For the ContextServ platform to access context sources, \textit{context providers} must be registered in the platform. Currently the platform supports two types of context providers: local context providers and remote context providers. Local context providers are responsible for collecting local context information, such as device memory capacity, CPU usage etc. On the other hand, remote context providers gather context information from a remote sensor or device. Every context provider has at least one \textit{agent}---a piece of program specifying the protocol on how to access the context information. 

%
%

\begin{figure*}[t]
\centering
\includegraphics[width=\linewidth]{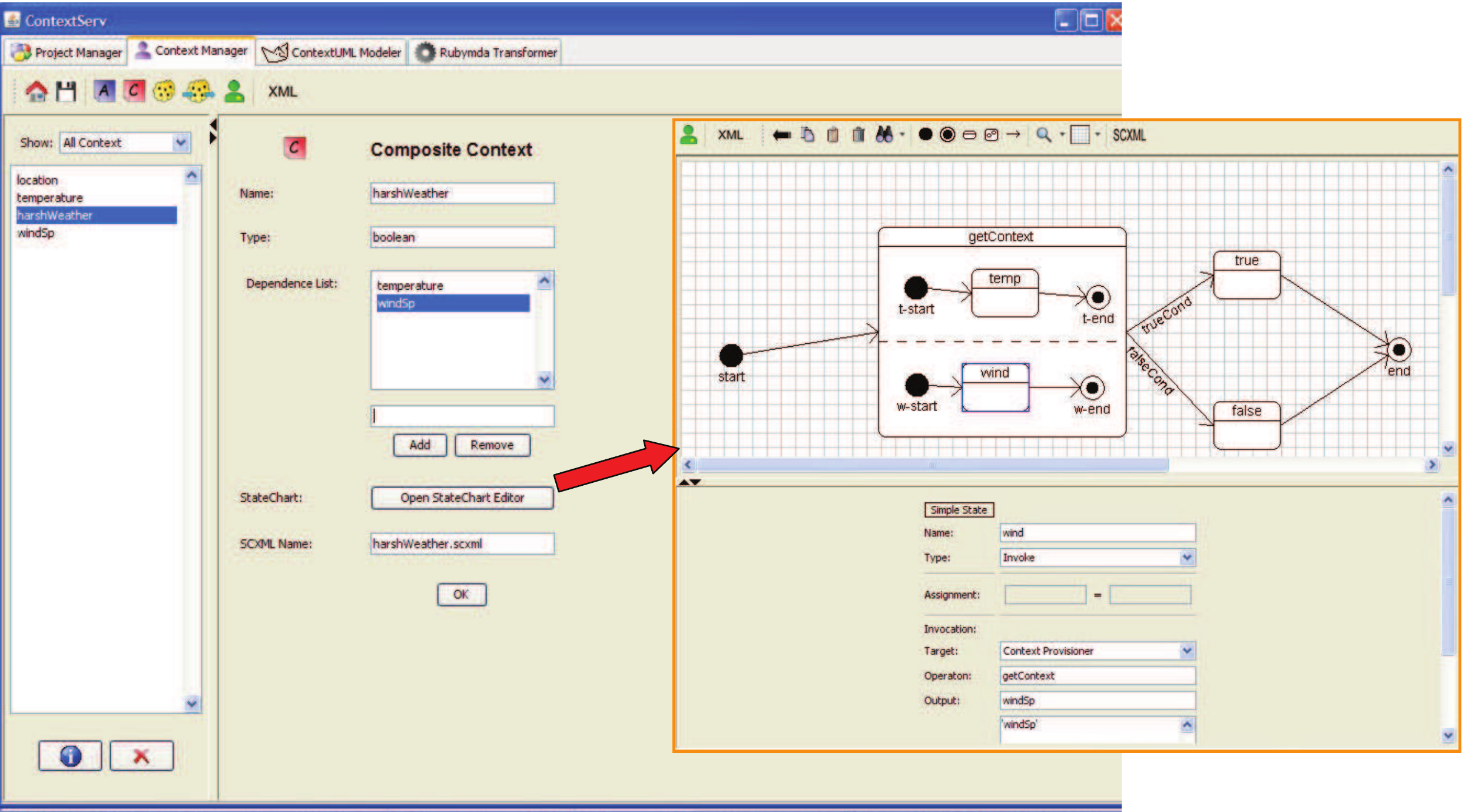}
\caption{Specifying composite contexts}
\label{fig:context}
\end{figure*}

\begin{figure*}[t]
\centering
\includegraphics[width=\linewidth]{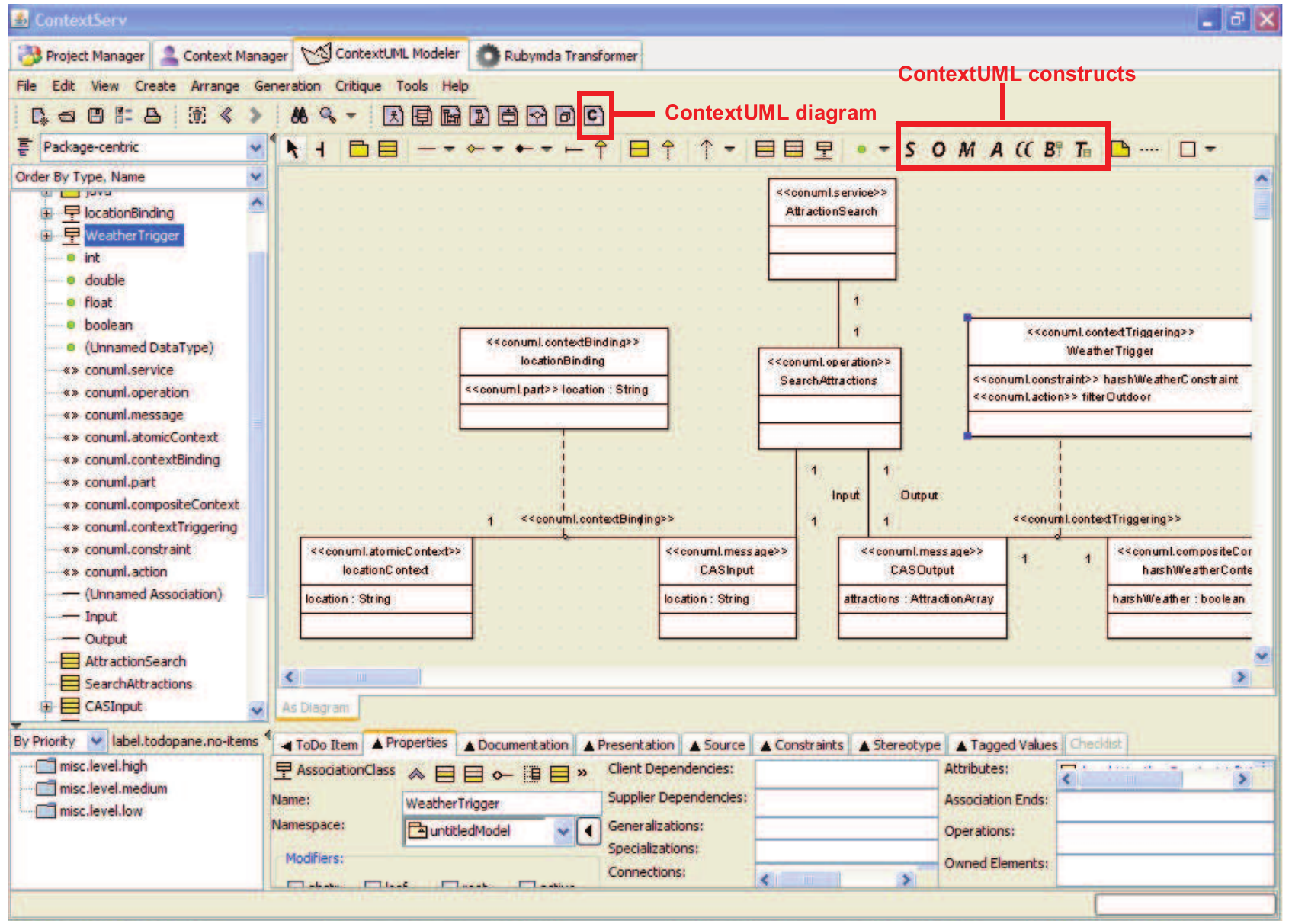}
\caption{The ContextUML Modeler}
\label{fig:contextUML}
\end{figure*}

Composite contexts are modelled using statecharts in ContextServ,
as shown in Figure~\ref{fig:context}. Statecharts is a widely used formalism that is emerging as a standard for process modelling following its integration into UML.
The statechart of a composite context is then exported into State Chart Extensible Markup Language (SCXML) \cite{Ayed2007MDD} 
an XML based language for describing generic statecharts, and executed in a SCXML execution engine such as Commons SCXML\footnote{http://commons.apache.org/scxml.}. 

\subsubsection{Managing Context Service Community}
A context service community implements a common interface (\texttt{add\-ContextSource()}, \texttt{removeConte\-xtSource()}, \texttt{selectContext\-Source()}) for context sources that provide the same context information. The main purpose of a context service community is to ensure robust and optimal provisioning of contexts to the context consumer so that on one hand a candidate context source can take the place of an unavailable context source, and on the other hand contexts having the best quality can be provisioned. More details on the implementation of context service communities will be reported in Section~\ref{sec:community1}.

\subsection{ContextUML Modeller}
The ContextUML modeler provides a visual interface (Figure~\ref{fig:contextUML}) 
for defining context-aware Web services using ContextUML. 
In the implementation, we extended ArgoUML\footnote{http://argouml.tigris.org.}, an existing UML editing tool, 
by developing a new diagram type, ContextUML diagram, which implements all the abstract syntax of the ContextUML language~\cite{Sheng-ICMB05}.

\subsection{RubyMDA Transformer}
\label{sec:Transformer}
Services represented in ContextUML diagrams are exported as XMI files for subsequent processing by 
the RubyMDA transformer, which 
is responsible for transforming ContextUML diagrams into executable Web services, using Ruby\-Gems\footnote{https://rubygems.org/.}. 
The ContextServ platform currently supports WS-BPEL, a de facto standard for specifying executable processes. 
Once the BPEL specification is generated, the model transformer deploys the BPEL process to an application server and exposes it as a Web service. In the implementation, JBoss Application Server is used since it is open source and includes a 
BPEL execution engine jBPM-BPEL.

\begin{table}[b] 
\centering
\caption{RubyMDA's Model Transformation Rules}
\begin{small}
\begin{tabular}{|p{7.0cm}|p{4.5cm}|}
\hline
\textbf{Map From: UML Stereotypes} & \textbf{To: BPEL Elements}  \\ \hline
 conuml.service &  \textless process\textgreater  \\ \hline 
 conuml.operation  &  \textless invoke\textgreater \\ \hline
 conuml.message & \textless variable\textgreater  \\ \hline 
 conuml.atomicContext & \textless invoke\textgreater  \\ \hline 
 conuml.compositeContext &  \textless invoke\textgreater    \\ \hline 
 conuml.contextBinding & \textless assign\textgreater   \\ \hline 
 conuml.part & \texttt{part} attribute in \textless to\textgreater  \\ \hline 
 conuml.contextTriggering & \textless switch\textgreater \textless invoke\textgreater  \\ \hline 
\end{tabular}
\end{small}
\label{tab:mapping}
\end{table}

RubyMDA is developed based on the model transformation rules. The model transformation rules are mappings from ContextUML stereotypes to BPEL elements. Table~\ref{tab:mapping} shows a summary of the model transformation rules of Ruby\-MDA. 

\begin{figure}[!htb]
	\centering
\includegraphics[width=0.7\linewidth]{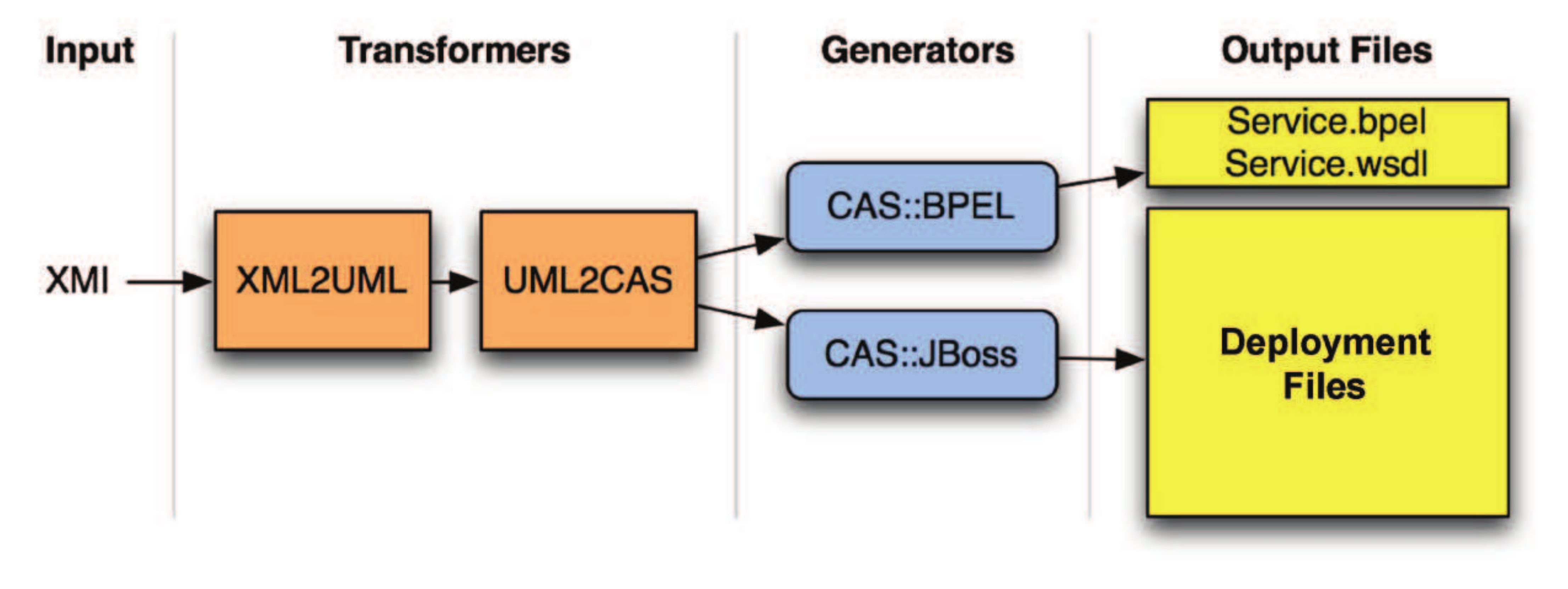}
	\caption{RubyMDA data flow}
	\label{fig:dataflow}
\end{figure}	

Figure~\ref{fig:dataflow} shows the data flow of RubyMDA model transformer. RubyMDA takes the XMI document as an input which represents the ContextUML diagram. RubyMDA reads the XMI document and constructs the UML model which is a set of data structure representing the components in UML class diagram. After the UML model is constructed, RubyMDA transforms it into CAS model which is a set of data structure representing the CAS described in ContextUML diagram. Finally, RubyMDA generates a BPEL process and WSDL document for a CAS. Moreover, it generates a set of deployment files needed to deploy CAS to a server.

\section{Dynamically Adaptive Process}
\label{sec:MoDAR}
Dynamic adaptability, which requires an application to adapt to new context at runtime, is a key challenge in developing composite web services, including CASs. As stated in~\cite{Papazoglou07}, ``{\em services and processes should equip themselves with adaptive service capabilities so that they can continually morph themselves to respond to environmental demands and changes without compromising operational and financial efficiencies}''. It is particularly important to CASs to cope with the functional changes raised from both business requirements and environmental contexts, and bringing dynamic adaptability(or agility) to service processes, which means a CAS should have the ability of behavior adaptation.

In \cite{yu2015model}, we proposed an approach called MoDAR (\underline{Mo}del-Driven Development of \underline{D}ynamically Adaptive Service-Oriented Systems with \underline{A}spects and \underline{R}ules) to support the systematic development of dynamically adaptive BPEL-based service-oriented systems. The integration of MoDAR and ContextServ can provide effective and efficient supports to the development of CASs with runtime behaviour adaptation. 
In this section, we introduce the MoDAR approach (Section \ref{sec:modar overview} to Section \ref{sec:MoDAR platform}) and how it enables dynamically adaptive CASs in ContextServ (Section \ref{sec:EnablingDACAS}). 

\begin{figure}[b]
	\centering
	\includegraphics[width=0.75\linewidth]{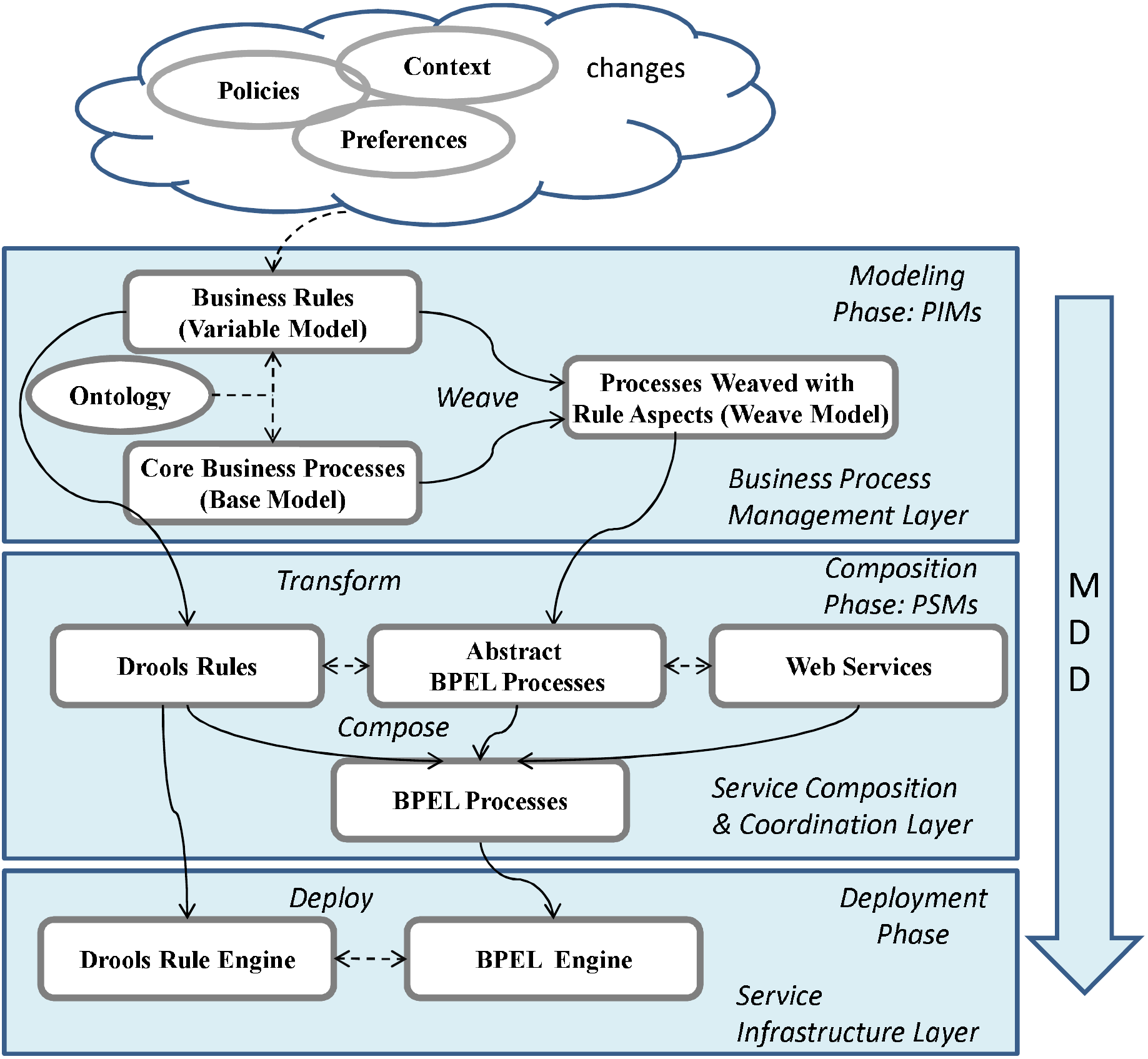}
	\caption{Overview of the MoDAR approach}
	\label{fig:modar_overview}
\end{figure}

\subsection{Overview of MoDAR}
\label{sec:modar overview}

As illustrated in Figure~\ref{fig:modar_overview}, the MoDAR approach consists of three main phases: i) the {\em modeling} phase, ii) the {\em composition} phase, and iii) the {\em deployment} phase.

MoDAR PIMs which include the {\em base model}, the {\em variable model}, and the {\em weave model}, are created in the modeling phases. We adopt the {\em separation of concerns} principle~\cite{Yasushi-Software2004} to manage complexity and variability: a system model is divided into a base model and a variable model. The base model represents the relatively stable processing procedures, or {\em flow} logic, of the system; while the variable model represents the more volatile {\em decision} aspect of the business requirements. To make the base model and the variable model semantically inter-operable, we use a minimum set of ontology concepts as the basic elements in defining activity parameters in processes and also in defining rule entities. We also adopt an aspect-oriented approach to integrate the base model and the variable model using a {\em weave model}. This approach ensures the modularity of the base model and the variable model so that they can evolve  independently~\cite{cibran2003aspect}. The MoDAR Models will be detailed in Section~\ref{sec:MoDAR models}.  

In the composition phase, the variable model is automatically transformed into Drools rules, and the weave model is automatically transformed into an abstract BPEL process. In this abstract process, at every {\em join point}, the invocation to a rule aspect is translated to a special Web service invocation. After the designer manually associates concrete Web services with abstract services in the process to implement their functionalities, the process is automatically transformed into an executable BPEL process.
In the deployment phase, the BPEL process and the Drools rules are deployed to their corresponding engines. Dynamic adaptivity is achieved in a way that 
we can freely add/remove/replace business rules defined in the modeling phase and then transform and redeploy them without terminating the execution of the process. 

\subsection{The MoDAR Models}
\label{sec:MoDAR models}
\subsubsection{The Base Model}
\label{sec:base_model}
The base model is used to capture the flow logic, or procedures, defined in the requirements. To promote the notion of treating procedures and decisions as two independent elements in a requirement, we use a subset of a full-fledged workflow language as the language for specifying the base model, and in this subset, only procedural elements are included while excluding all the decision elements. 

To avoid creating yet ``another" workflow language, we reuse the flow logic related elements defined in BPMN, while extending the original {\it Business Activity} element with properties for associating with the variable model. In general, the base model language has two types of key elements: the {\it Flow Object} and the {\it Connecting Object}, where flow objects are processing entities and connecting objects specify the flow relations between flow objects. 
There are three types of flow objects: the {\it Business Activity}, the {\it Event}, and the {\it Parallel Gateway}. Business activities represent the main processing unit of a requirement. A business activity can be defined as {\it variable} to indicate that this activity is adaptive at runtime. Events have the normal meaning as defined in BPMN: they are happenings that affect the execution of a process (e.g.,exceptions). To promote the separation between procedures and decisions, we only include parallel gateways---which represent the forking of process flow---in the base model while excluding all the other decision-related BPMN gateways. All the decision-related requirements will be captured by the MoDAR variable model. 

To create a base model, we can start from scratch to describe the general steps of a business process and then connect these steps using sequential or parallel flows. Alternatively, we can re-engineer existing business processes and derive base models from them.

\subsubsection{The Variable Model}
\label{sec:var_model}
The variable model is used to capture the decision aspect of a business requirement, which is changeable at runtime. 
A variable model consists of a set of business rules where each rule $r$ is defined by a 3-tuple: $r$ = $<$$type, condition, action$$>$. A rule type can be either $constraint$, or $computation$, or $inference$, or $action$. Our rule definition follows the typical event-condition-action (ECA) pattern but the {\it event} part is specified in the weave model (see the next subsection for details) because the triggering of rules is determined by point cuts in the aspect.

we have designed a high-level rule language and its associated graphical editing tool to facilitate the specification of business rules. The syntax of this rule language is defined as follows:
\begin{footnotesize}
\begin{lstlisting}
<rule>     ::= <type>, <cond>, <action> 

<cond>     ::= not <cond> | <cond> and <cond> | 
               <cond> or <cond> | <term> <relop> <term>
<term>     ::= <property> | <term> <arop> <term> | 
               <const> | <fun> (<term> <term>*)
<property> ::= <concept>(_<n>)?(.<obj_prop>)*.
			   <datatype_prop>
<relop>    ::= less than | less than or equal to | 
               equal to | greater than or equal to | greater than
<arop>     ::= + | - | * | / 
<n>        ::= 1 | 2 | 3 |... 

<fun>      ::= <predef> | <usrdef>
<predef>   ::= abs | replace | substring | sum | avg 
               | min | max | ...

<action>   ::= (<act>) | (<property> | <concept>(_<n>)? 
               = <term> | <act>))* 
<act>      ::= <activity> | Skip <activity> | 
               Skip <activity> Then <activity> | Abort | 
               <activity> Then Abort
               

\end{lstlisting}
\end{footnotesize}

There are two main features in this rule language: i) ontology concepts and properties are introduced in the specification of both the {\it condition} and the {\it action} of a rule; ii) user defined functions, such as the {\texttt distance} function for calculating the distance between two places, are allowed in defining the condition of a rule. The benefit of the second feature is straightforward---a domain specific function library can be built to facilitate the definition of complex rules. 


Based on Web Ontology Language (OWL)~\cite{Mcguiness2004}, an ontology concept could be a complex structure having both {\it object properties} and {\it datatype properties}, where an object property navigates to another concept in the ontology and a datatype property has a specific primitive data type such as {\em integer}, {\em boolean}, or {\em string}. 
For example, suppose the {\texttt Customer} concept has an object property {\texttt contact} whose range is the concept {\texttt Contact}, and {\texttt phoneNumber} is a {\em string} datatype property of {\texttt Contact}. Only datatype properties are allowed in defining condition terms because operations on objects are not defined in the context. 
For the action part, we can either assign the result of a term expression to a variable, or assign the result of the invocation of a business activity to a variable.

\subsubsection{The Weave Model}
\label{sec:weave-model}

The decision logic defined in the variable model is associated to the base model by the weave model, which is composed of a set of aspects.
Each aspect $a$ weaves a rule set to a business activity in the base model: $a \in \{Before, Around, After\} \times \mathcal T \times  2^\mathcal R$, where $\mathcal T$ is the set of business activities and 
$2^\mathcal R$ is the set of rule sets. 

Similar to AspectJ, an aspect-oriented programming (AOP) extension created at PARC for the Java programming language, we also identify three types of aspect: {\it before aspects}, {\it around aspects}, and {\it after aspects}. An aspect is always associated with a business activity. Both before aspects and around aspects are activated before the execution of the associated activity, but if an activity has an 
around aspect, this activity will not be executed after the around aspect. From the perspective of the ECA pattern, $event \in \{Before, Around, After\} \times \mathcal T$ becomes the triggering event of a rule. It is worth noting that the interoperability between an activity and its associated rules 
are established through the predefined ontology.

\subsection{The MoDAR Platform}
\label{sec:MoDAR platform}
The MoDAR development environment has two main components: i) the {\it Process Modeler} for graphically modeling the base model, the rules, and the weave model, and ii) the 
{\it Association and Transformation Tool} for associating Web services with activities defined in the base model and for generating and deploying executable code. To facilitate process and rule definition, the development environment also has a {\it Business Domain Explorer} component for graphical exploration of domain ontologies defined in OWL (the OWL files are created using ontology editing tools such as Prot\'{e}g\'{e}\footnote{http://protege.stanford.edu/}).
The Process Modeler 
offers a visual interface for defining the base model, the variable model, and also the weave model.
The association and transformation tool has two main functionalities. First, it provides a visual interface for associating Web services to activities defined in both the base model (the BPEL process) and the variable model (within the $action$ part of rules). Second, it is used to automatically generate executable code and deployment scripts from the models defined in the modeling phase. In the current implementation BPEL was selected as the targeted executable process language and Drools as the targeted executable rule language.

\begin{figure}[!tb]
	\centering
	\includegraphics[width=0.8\linewidth]{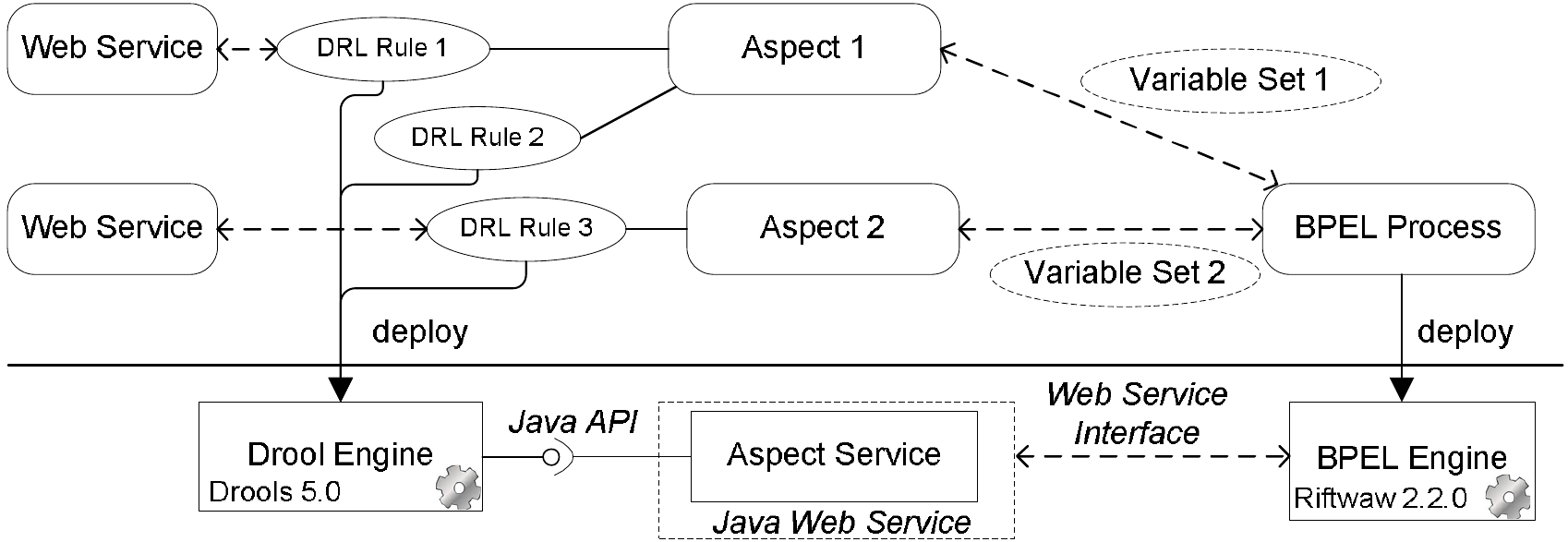}
	\caption{An anatomy of the MoDAR runtime environment}
	\label{fig:runtime}
\end{figure}

Figure~\ref{fig:runtime} is an anatomy of the MoDAR runtime environment. The bottom level of the anatomy includes the main components of the runtime environment: a {\it Drools engine (Drools)}, a {\it  BPEL engine (Riftsaw)}, and a {\it  generic aspect service} that encapsulates the rule invocation logic. 
The aspect service is written in Java and exposed as a Web service for the BPEL process to invoke. Every time when an aspect in the process is reached for execution, the aspect service is invoked and corresponding variables (including the IO parameters of its associated activity and user selected variables) are passed from the process to it; these variables are used in the execution of the rules of the aspect. After all the rules in the aspect are executed, these variables are updated and passed back to the process. As discussed in the above subsection, a Drool rule can invoke external Web services to implement action-enabler business rules. 
In case an exception is raised when executing a Drool rule or invoking an external Web service, the exception will be propagated by the aspect service to the base process for exception handling. For error message traceability between the Drools engine and the BPEL engine, our current implementation relies on the built-in logging functionality of Drools\footnote{http://docs.jboss.org/jbpm/v5.2/javadocs/org/drools/logger/KnowledgeRuntimeLoggerFactory.html}.
A further extension to the aspect service would be passing the error message or error code generated by the Drools engine directly to the BPEL engine.

\subsection{Enabling Dynamically Adaptive CASs}
\label{sec:EnablingDACAS}
To enable the dynamic adaptivity feature of CASs developed by the ContextServ platform, the {\small \sf ContextTriggering} elements specified as WS-BPEL {\small \sf \textless switch \textgreater} and  {\small \sf \textless invoke \textgreater} (cf. Section~\ref{sec:Transformer}) are first automatically extracted from the original CAS WS-BPEL script, and then transformed into MoDAR rules, and finally weaved into the WS-BPEL process as an \emph{after aspect}. 

For example, the weather condition based service filtering context trigger discussed in Section~\ref{sec:binding} is transformed into the following MoDAR Rule:

\begin{quotation}
\noindent
{\it $\mathcal R_1$: If temperature is greater than 30 degree, and wind speed is greater than 25 km/s, filter out outdoor activities:} 
\begin{footnotesize}

\begin{lstlisting}
[Cond]     Weather.temperature greater than "30" and 
           Weather.windspeed greater than "25"
[Action]   Filter("Filter out outdoor 
           activities", ActivityList).
\end{lstlisting}

\end{footnotesize}
\end{quotation}

This rule is then weaved into the original process \emph{after} the generic attraction finder services.

At runtime, if the developer or user wants to change the context-aware triggering condition, such as adjusting the temperature and/or wind speed threshold, or adding a new condition such as \emph{raining}, then the user can just change the rule on-the-fly without stopping the execution of the composite CAS.




\section{Optimization of Context Service Community}
\label{sec:community1}

As previously mentioned, \textit{context service community}(CSC for short) is an important concept for optimized and dynamic context information provision, which directly relates to the quality of context-aware Web services. In the ContextServ platform, we develop a module called 
context service community manager (CSCM) for managing CSCs in ContextServ.   
In this section, we describe the technical details of CSCM.

CSCM aims at solving the following two issues:

\begin{itemize}
\item A single context service may fail to provide the requested context information due to various reasons such as the server is unavailable or the associated sensor is broken down.

\item There might exist multiple context services providing a same piece of context information (usually with different levels of quality). It is hard for a CAS to ensure that the context service it interacts with provides the optimal context information.
\end{itemize}

CSCM serves as a broker between a CAS and its context providing services and will select the context service that provides optimal context for the CAS from all the context services. CSCM has four main components: 

\begin{itemize}
\item The \textit{Context Retrieval Process} is implemented as a manager for retrieving and unifying contexts from heterogeneous sources;
\item The \textit{Context Monitoring Process} is responsible for actively monitoring the quality information of the registered context services by collecting and keeping their quality information obtained through the context retrieval process;
\item The \textit{Context Evaluation Process} is responsible for evaluating the quality information obtained from the context monitoring process;
\item The \textit{Context Service Selection Process} is implemented for selecting the best context service that provides the optimal context information at runtime by using the results obtained from the context evaluation process. 
\end{itemize}

Besides these four components, CSCM also has two user interface components for context service providers and context service consumers respectively. Through the user interface for context servcie providers, context servcie providers can register to become a member of a CSC for their context information to be used by supplying required registration details such as the execution prices of their services. Also, a CSC can advertise its required context information from demands of context service consumers to context servcie providers. The user interface for context service consumers is similar to the one for context service providers.

\subsection{Quality Parameters for Context Service Provider}
\label{sec:Quality Parameters}

In a CSC, in order to select the best context service that provides the optimal context information, it is important to consider both the quality of context information provided by the context service (i.e. QoC) and the quality of context service provided by the context service provider for getting the context information (i.e. QoS). Hence the quality of context service provider is modelled by a set of QoC and QoS quality attributes. 

Many QoC and QoS attributes have been proposed and defined~\cite{QoC-03,kim2006QoC,nazario2014QoC,al2007qos,liu2004qos,zeng2003quality,SELFSERV-DAPD}. Decision of what and how many specific quality attributes should be defined for CSC is hard to make due to the reason that requirements from CAS customers are always different. Furthermore, both QoC and QoS attributes need to be clearly identified. Based on our best experience and a review of the related literature, so far, we consider nine generic quality attributes for CSC which consist of five QoC attributes (\textit{precision, trustWorthiness, correctnessProbablity, refreshRate} and \textit{up-to-dateness}) and four QoS attributes (\textit{executionPrice, responseTime, availability} and \textit{reliability}) as shown in Table~\ref{tab:QualParameters}.

\begin{table}[htbp]
	\centering
	\caption{RubyMDA's Model Transformation Rules}
        \begin{small}
		\begin{tabular}{|p{2.6cm}|p{7.0cm}|p{1.3cm}|p{1.5cm}|}
		
			\hline
			\textbf{Attribute} & \textbf{Aggregation Function} & \textbf{Category} & \textbf{Source}\\ 
			\hline
			precision & $Q_{pre}(cs)$=\textit{registered value}  &  QoC &  provider \\ 
			\hline 
			trustWorthiness  &  $Q_{truW}(cs)=\frac{1}{H}\sum_{k=1}^Htw_k(cs)$ &  QoC &  community \\ 
			\hline
			correctnessProbablity & $Q_{corP}(cs)$=\textit{registered value}   &  QoC &  provider \\ 
			\hline 
			refreshRate & $Q_{refR}(cs)$=\textit{registered value}  &  QoC &  provider \\ 
			\hline 
			up-to-dateness &  $Q_{utd}(cs)=1-(t_{cur}-t_{med})/t_{\theta}$ if  $t_{cur}-t_{med}<t_{\theta}$; 0 \textit{otherwise}  &  QoC &  community \\ 
			\hline 
			executionPrice & $Q_{price}(cs)$=\textit{registered value}  & QoS &  provider \\ 
			\hline 
			responseTime &  $Q_{resT}(cs)=\frac{1}{H}\sum_{k=1}^Hrs_k(cs)$  &  QoS &  community\\ 
			\hline 
			availability & $Q_{ava}(cs)=T_{a}(cs)/\theta$ &  QoS &  community\\ 
			\hline 
			reliability & $Q_{rel}(cs)=\frac{1}{H}\sum_{k=1}^Hre_k(cs)$  &  QoS &  community \\ 
			\hline 
			
		\end{tabular}
		\end{small}
	\label{tab:QualParameters}
\end{table}

\begin{itemize}
	\item \textbf{\textit{precision}}. It represents how accurate the context information provided by context service provider is and how exactly the provided context information mirrors the reality.
	
	\item \textbf{\textit{correctnessProbablity}}. It refers to the probability that the provided context is correct.
	
	\item \textbf{\textit{refreshRate}}. It indicates the rate that the context information provided by context service provider is updated. 
	
	\item \textbf{\textit{ executionPrice}}. It represents the price needed to pay for executing a service provided by context service provider for getting the context information.
\end{itemize}

In a CSC, initial values of the above four quality attributes are set to the registered values provided by context service providers. A CSC does not involve in how to calculate the values of the four attributes. 

With regard to the other five attributes, their initial values are calculated from the log files maintained by the Context Monitoring Process of CSCM, which are in XML format and store the history information of context service providers. 

\begin{itemize}
	\item \textbf{\textit{trustworthiness}}. The trustworthiness $Q_{truW}(cs)$ of a context service $cs$ represents how much the context information provided by context provider is trusted by context consumer which mainly depends on end consumer's experiences. Different end consumers may have different opinions on the same service. The value of the trustworthiness is defined as the average ranking given to the service by end consumers, i.e., $Q_{truW}(cs)=\frac{1}{H}\sum_{k=1}^Htw_k(cs)$, where $tw_k(cs)$ is the end consumer's ranking on $cs$, $H$ is the number of times the service has been graded. The ranking has a range $[0,5]$. 

	\item \textbf{\textit{up-to-dateness}}. It indicates how old the context information is by using a timestamp. If the current time $t_{cur}$ minus the information measured time $t_{med}$ is less than $t_{\theta}$, which is the context information lifetime, the up-to-dateness $Q_{utd}(cs)$ equals to $1-(t_{cur}-t_{med})/t_{\theta}$, otherwise $Q_{utd}(cs)=0$ which means the information is outdated. The larger the value
	of $Q_{utd}(cs)$ is, the newer and better quality the context information is.
	
	\item \textbf{\textit{responseTime}}. The response time $Q_{resT}(cs)$ of a service $cs$ refers to the time taken from sending a request to receiving a response from $cs$ in order to obtain the context information. In CSC, the value of response time is defined as the average response time in historical data about past invocations using the expression $Q_{resT}(cs)=\frac{1}{H}\sum_{k=1}^Hrs_k(cs)$, where $rs_k(cs)$ is the response time of $cs$ in the $kth$ invocation, $H$ is the total number of invocations.
	
	\item \textbf{\textit{availability}}. The availability $Q_{ava}(cs)$ of a context service $cs$ is the probability that the service is accessible. The value of the availability of a context service $cs$ is computed using the following expression $Q_{ava}(cs)=T_a(cs)/\theta$, where $T_a$ is the total amount of time (in seconds) in which service $cs$ is available during the last $\theta$ seconds ($\theta$ is a constant set by an administrator of CSC). The value of $\theta$ may vary depending on a particular context service. For example, if a context service is more frequently accessed (e.g., location), a small value of $\theta$ gives a more accurate approximation for the availability of services. If the context service is less frequently accessed (e.g., weather), using a larger $\theta$ is more appropriate. Here, we assume that context services send notifications to CSC about their running states (i.e., available, unavailable).
	
	\item \textbf{\textit{reliability}}. The reliability $Q_{rel}(cs)$ of a context service $cs$ is the probability that a request is correctly responded within the maximum expected time frame. In CSC, the value of reliability is computed from historical data (stored in log files) about past invocations using the expression  $Q_{rel}(cs)=\frac{1}{H}\sum_{k=1}^Hre_k(cs)$, where  $\sum_{k=1}^Hre_k(cs)$ is the number of times that the service $cs$ has been successfully delivered within the maximum expected time frame, and $H$ is the total number of invocations. The value of $re_k(cs)$ is 1 if service is successfully executed, 0 otherwise.
	
\end{itemize}

\subsection{Optimal Context Service Selection}
\label{sec:selection1}
Suppose that a specific piece of context $c$ (e.g. \emph{temperature}) can be supplied by a set of context service providers i.e., $\left\{\mathcal {CSP}_{i}\right\}_{1\leq i \leq n}$. These context service providers are members of a CSC (e.g., a weather CSC $\mathcal CSC_{w}$). 
The quality of context service provider for each $\mathcal{CSP}_{i}$ is modelled by a set of $m$ quality attributes $\left\{\mathcal A_{j}\right\}_{1\leq j \leq m}$, e.g. precision, response time, availability, and a distribution of weights $\left\{\mathcal{W}_{j}\right\}_{1\leq j \leq m}$ over these attributes where $\sum_{j=1}^m w_{j}=1$. 

To calculate the score of each context service provider, we use the following multi-attribute utility function~\cite{SELFSERV-DAPD}:
	\begin{equation} \label{eq:rp}
			U(\mathcal{CSP}_{i}) = \sum_{j=1}^m \mathcal{W}_{j} \times \mathcal S_{i,j}
	\end{equation}

\noindent where $\mathcal S_{i,j}$ represents the score of $\mathcal A_{j}$ of $\mathcal{CSP}_{i}$.

The score matrix $\mathcal S$ is derived from scaling the initial attributes value matrix $\mathcal I$. 
Since there are \textit{positive} attributes (e.g., avalilability) where a greater value indicates a better quality, and also \textit{negative} attributes (e.g., refresh rate) where a less value indicates a better quality, the score of each $\mathcal A_j$ is calculated differently for positive attributes and negative attributes:

	\noindent  If $\mathcal A_{j}$ is negative
	\begin{equation}
			\mathcal S_{i,j} = \left\{ \begin{array}{ll}
			\frac{\mathcal I^{max}_j - \mathcal I_{i,j}}{\mathcal I^{diff}_j} & \textrm{$\mathcal I^{diff}_j  \neq 0$}\\
			1 & \textrm{$\mathcal I^{diff}_j = 0$}\\
			\end{array} \right.
	  \end{equation}
	  If $\mathcal A_{j}$ is positive
	  \begin{equation}
			\mathcal S_{i,j} = \left\{ \begin{array}{ll}
			\frac{\mathcal I_{i,j} - \mathcal I^{min}_j}{\mathcal I^{diff}_j} & \textrm{$\mathcal I^{diff}_j  \neq  0$}\\
			1 & \textrm{$\mathcal I^{diff}_j = 0$}\\
			\end{array} \right.
	  \end{equation}
		Where: $
			\mathcal I^{max}_j = max(\mathcal I_{j}), \ \mathcal I^{min}_j = min(\mathcal I_{j}), \ \mathcal I^{diff}_j = \mathcal I^{max}_j - \mathcal I^{min}_j $

Based on the utility values, the ranking list of all context service providers can be produced. Theoretically, $\mathcal{CSP}_{i}$ with the highest utility value in the ranking list will be selected as the optimal context service provider for providing the context. However, in order to cope with the dynamicity of context service providers in the real time as the situation the best $\mathcal{CSP}_{i}$ chosen is suddenly unavailable, the select process considers not only the ranking but also the availability in real time. Moreover, the select process is also extended to handle a context consumer's specific constraints (e.g., maximum \textit{execution price} less than \textit{PRICE}, maximum \textit{response time} less than \textit{TIME}). More details on modeling quality attributes and implementation of the context service community component can be found in ~\cite{Liao2009}. 

It also should be noted that \textit{ontologies} play a pivotal role in understanding the semantics of the context services. Context services described in ontologies possess explicit semantic representations, which makes the automatic selection of context services possible. The description of the detailed ontology model of context services is outside of the scope of this paper.	 

\section{Evaluation}
\label{sec:Eval}
In this section we evaluate the core components of ContextServ: {\em the ContextServ development environment}, {\em the Context Service Community middleware} and {\em the MoDAR execution environment}. For the ContextServ development environment, we have conducted a controlled usability test to validate our hypothesis that ContextServ improves the efficiency of developing CASs. Then we report our performance studies on the Context Service Community middleware and the MoDAR middleware.

\subsection{Experimental Evaluation of the ContextServ Development Environment}

Because ContextServ includes a model-driven development environment aiming at providing a generic yet effective approach for developing CASs, using ContextServ, the user is expected to on the one hand be able to develop CASs at a higher-level of abstraction without concerning the detail of programming, and on the other hand improve the efficiency in developing CASs compared to programming from scratch. 

We conducted a controlled experiment to study the performance of programmers in developing CASs with or without ContextServ. The aim of this experiment is to validate our hypothesis that ContextServ improves the efficiency of developing CASs. In the rest of this section, we first describe the design of the experiment--including the study subjects and the scenario used in the study, and then we demonstrate the result and analyze the result statistically.

\subsubsection{Design of the Experiment}
The design of the evaluation include two parts: study subject and scenarios used in the study. We will discuss them in this section. 
\\

\noindent
{\em Study Subjects.}
Considering that development performance may vary significantly between experienced and novice programmers, we divided the subjects into two groups: an expert group, and a novice group.
The expert group 
consists of two programmers who have more than five years of programming experience and also participated in the development of the ContextServ platform.  

The novice group consists of 15 year-two undergraduate students in software engineering and computer science. These novice programmers have less than three years of programming experience and have no prior experience in developing CASs, but all of them have studies the Introduction to Software Engineering course and have used UML and Java in small projects.  
\\

\noindent
{\em Scenarios Used in the Study.}
The participants were asked to implement a sample context-aware application called Smart Shopping Guide\footnote{http://hs.cs.adelaide.edu.au/shoppingguide/demo.html}
 : {\em Smart Shopping Guide is a context-aware web service which automatically provides the shop news based on the users' current location in shopping mall and their shopping preferences. This service can enhance your shopping experience by automatically fetching the nearest shops' news. Moreover, you can also filter out the shops which you are not interested.}

Each participant was asked to implement the application twice: using ContextServ in one case and using Java in the other. To prevent scenario learning possibly biasing the accuracy of the evaluation, in the novice group, eight participants were asked to use ContextServ in the first place and then Java to implement the application, while the other seven participants were asked to use Java in the first place; similarly in the expert group, one is asked to use ContextServ first and the other Java first.

The time that the participants used on developing the application was recorded. Especially when using the ContextServ, time spent on each of the stages--including context modeling, application modeling, and model transformation--is recorded separately. It is worth noting that a 15 minute tutorial on ContextServ is also counted into the total time of development.
\subsubsection{The Experimental Result and Analysis}
Table~\ref{tab:experiment} shows the time spent in developing the sample application using both ContextServ and Java, and also the time difference between the ContextServ and the Java scenarios. 

\begin{table}
\caption{Results of the development efficiency experiment}
\begin{small}
\begin{tabular}{|c|c|c|c|c|}
\hline
&{\bf Participant ID} &{\bf Dev Time (Min)} &{\bf Dev Time (Min)} &{\bf Time Differences} \\
 & &ContextServ &Java & (Min)\\
\hline

\multirow{15}{*}{Novice Group}
&1* & 155 & 300 & -145\\
&2*	&240	&420	&-180\\
&3*	&190	&285	&-95\\
&4*	&120	&135	&-15\\
&5*	&315	&310	&5\\
&6*	&250	&325	&-75\\
&7*	&285	&310	&-25\\
&8*	&290	&400	&-110\\
&9		&280	&420	&-140\\
&10	&280	&320	&-40\\
&11	&230	&315	&-85\\
&12	&190	&290	&-100\\
&13	&240	&310	&-70\\
&14	&185	&280	&-95\\
&15	&290	&510	&-220\\
\hline
\multirow{3}{*}{Expert Group} 
&	&	&	&\\
&16*	&110	&100	&5\\
&17	&90	&90	&0\\
\hline
\multicolumn{4}{l}{{\em * Participants who use ContextServ in the first place}}\\  

\end{tabular}
\end{small}
\label{tab:experiment}
\end{table}

According to the result table, it is quite obvious that the novice programmers used much less development time in ContextServ than in Java. We statistically tested this hypothesis by assessing the difference between the ContextServ scenario and the Java scenario with Student t-test~\cite{Montgomery2003}. The null hypothesis is:

$H_0:$ {\em For novice programmers, using ContextServ DOES NOT improve the efficiency of development compared to using Java:}
\\ \\
\centerline{ $\mu_{ContextServ} - \mu_{Java} >= 0.$}
\\ \\
The alternative hypothesis is:

$H_1:$ {\em For novice programmers, using ContextServ DOES improve the efficiency of development compared to using Java:}
\\ \\
\centerline{ $\mu_{ContextServ} - \mu_{Java} < 0.$}
\\ \\

\begin{figure}[]
	\centering
	\includegraphics[width=0.8\linewidth]{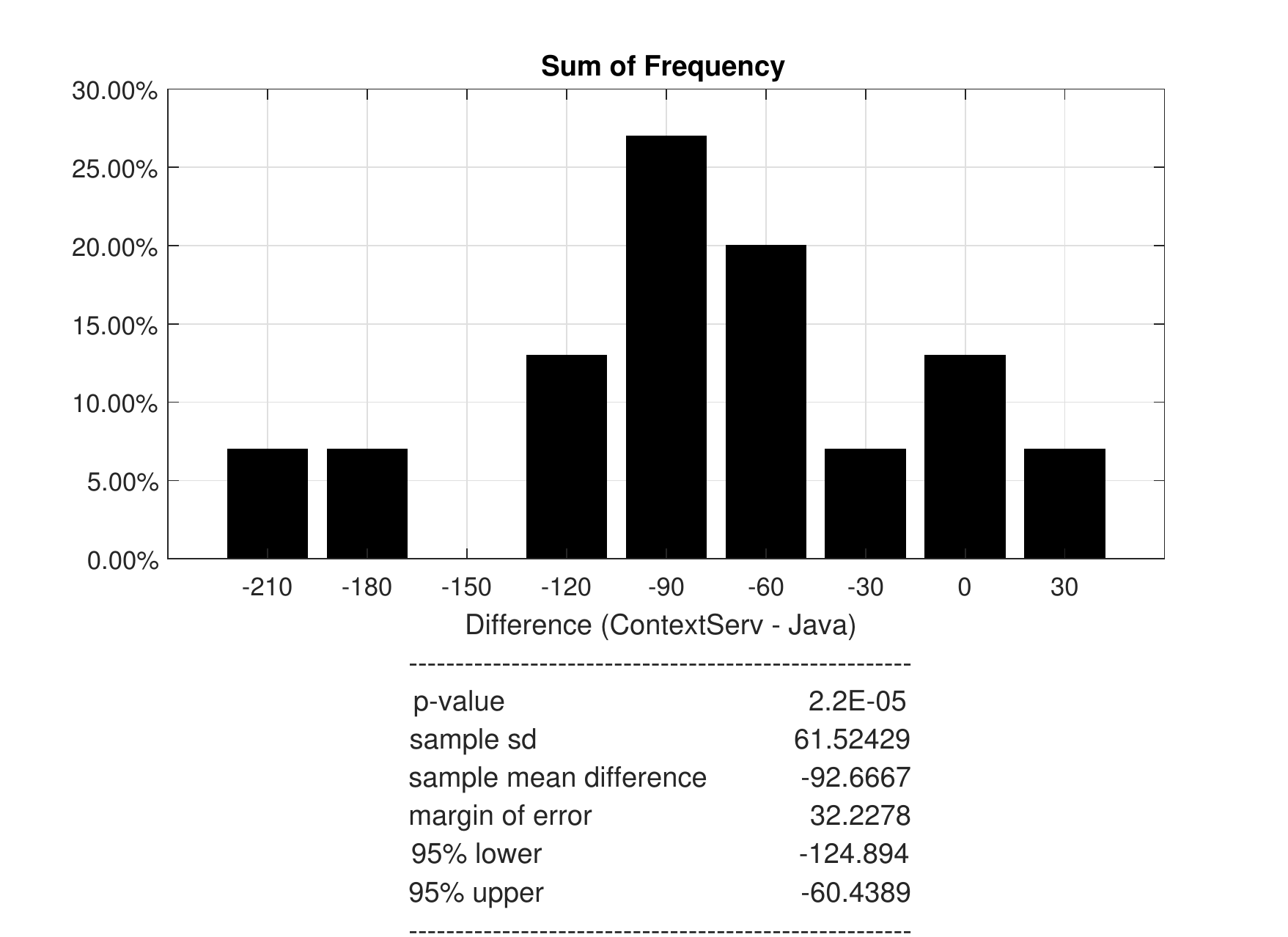}
	\caption{{\em t-test} result of the experiment}
	\label{fig:stat}
\end{figure} 

Figure~\ref{fig:stat} is the {\em t-test} result of the differences between development time of using ContextServ and using Java. The {\em p-value} is .00002. Were the ContextServ scenario does not improve the efficiency of development, it would be unusual to observe such a large sample mean difference. Based on sample evidence shown in Figure~\ref{fig:stat}, we can conclude that to novice programmers, ContextServ significantly improves their efficiency in developing CASs compared to programming from scratch. From Figure~\ref{fig:stat} we also know that for the Smart Shopping Guide scenario, ContextServ shortens the development time of novice programmers by about 60 to 125 minutes (on a level of significance of 5\%).

From the performance data of the two expert programmers, clearly they are more efficient developers than novice programmers. But there is very little difference in performance whether they use ContextServ or Java. 

We believe that supporting novice programmers has precedence because a user-centric development paradigm-in which inexperienced users are granted the freedom to compose their own applications-is gaining focus and momentum with the transformation of the Internet to a more open and personalized platform. The Web 2.0 phenomena also confirm that common business users will play an important role in developing personalized context-aware applications.

\subsection{Performance Study of the Context Service Community Middleware}
We have conducted several experiments to test the performance of the Context Service Community middleware. In particular, we conducted experiments to study i) the performance of our optimal context service selection algorithm presented in Section~\ref{sec:selection1},
and ii) the performance of the whole context selection process of Context Service Community. 
In the experiments, we constructed a weather context service community with 1,000 context service providers.

\begin{figure}
	\begin{minipage}[t]{0.5\linewidth}
		\centering
		\includegraphics[width=1\textwidth]{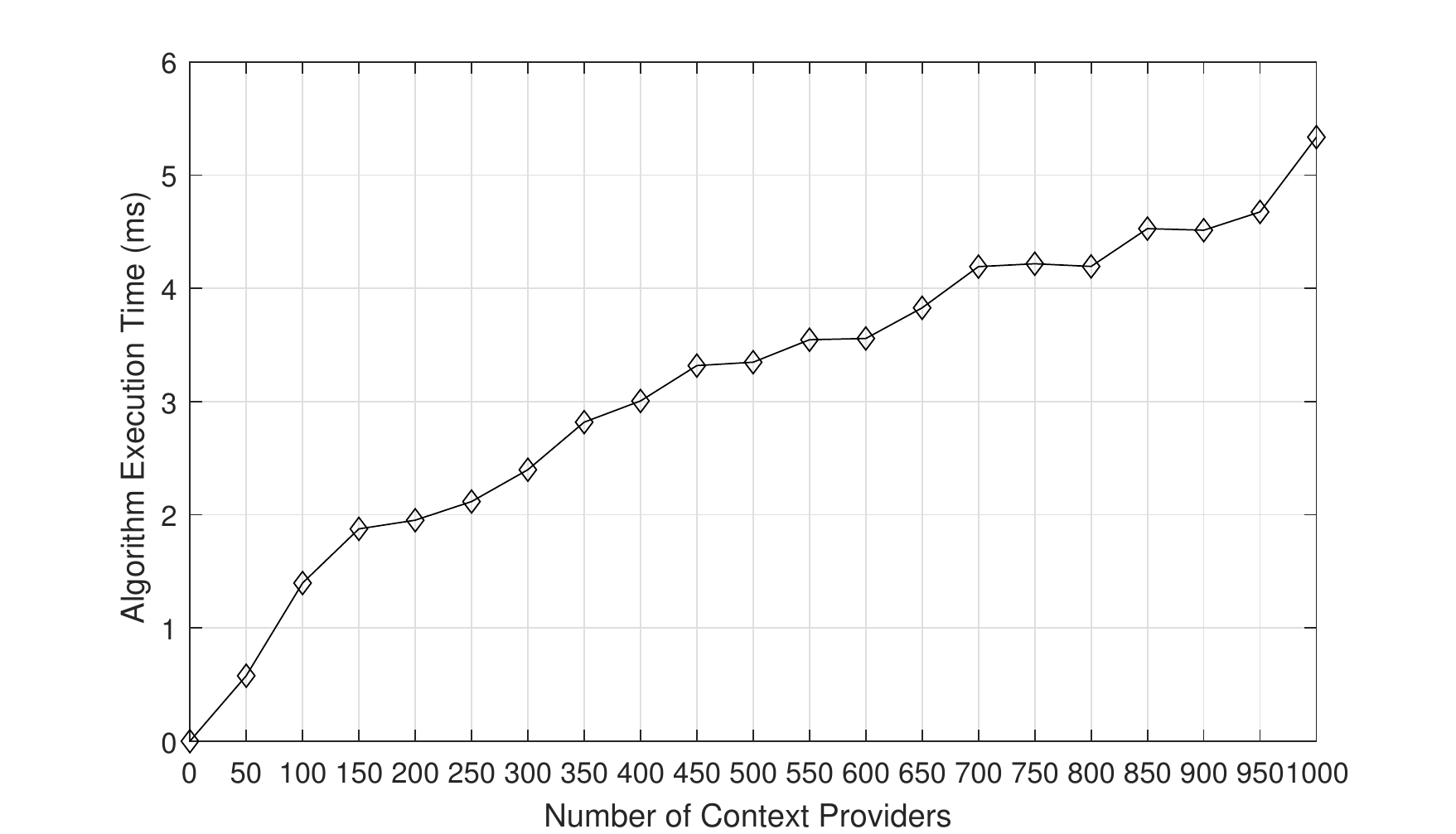}
		\caption{Performance of the selection algorithm}
		\label{fig:perf-algo}
	\end{minipage}%
	\begin{minipage}[t]{0.5\linewidth}
		\centering
		\includegraphics[width=1\textwidth]{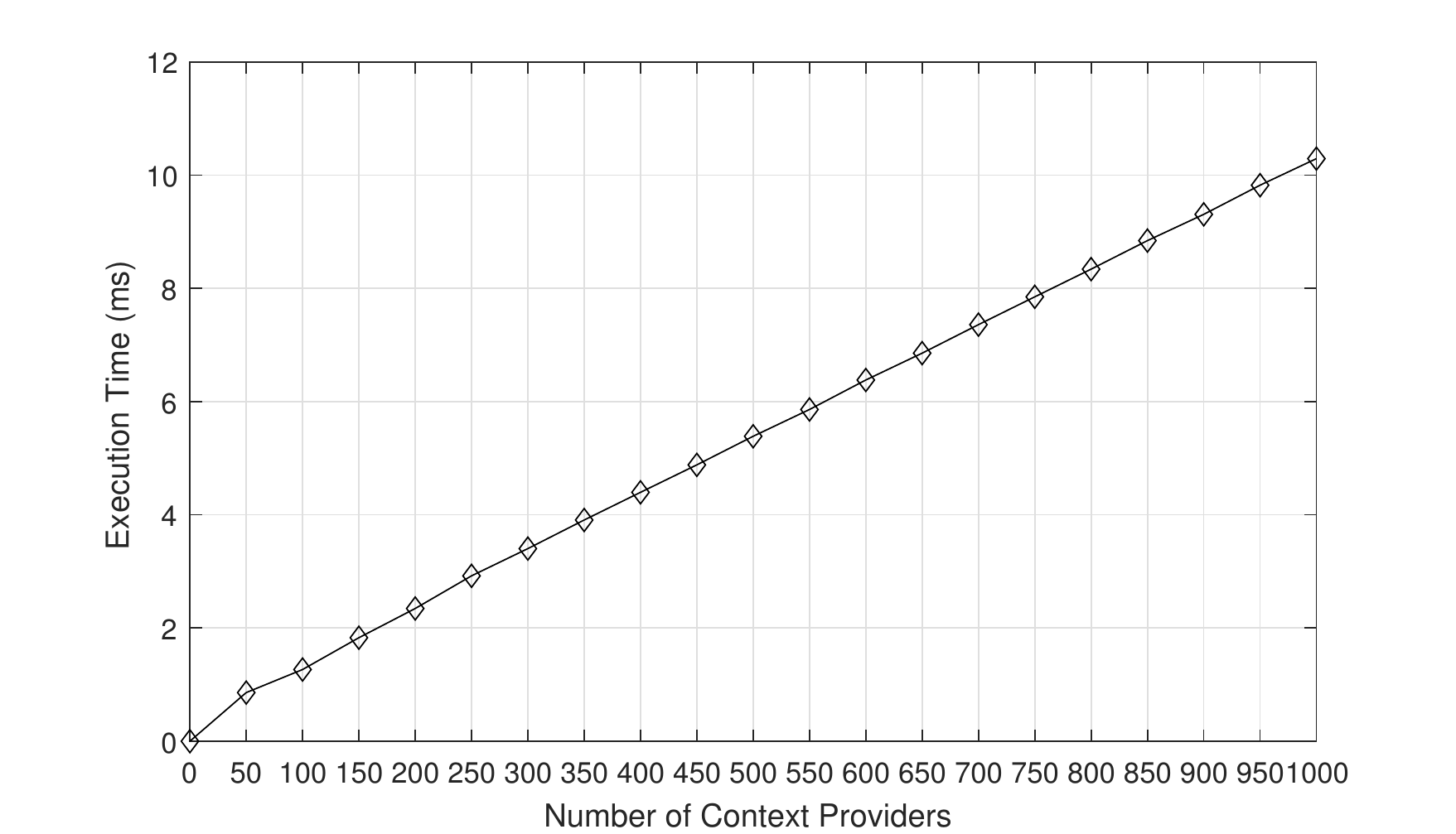}
		\caption{Performance of the full selection process}
	    \label{fig:perf-full}
	\end{minipage}
\end{figure}

In the first experiment, we evaluated the performance of our proposed context service selection algorithm. In this experiment, the initial value matrix $\mathcal I$ is already established in the context service community (i.e., we only evaluate the context service selection process, with no involvement of other three components such as the context retrieval process, the context monitoring process, and the context evaluation process, see Section~\ref{sec:community1}). We executed the selection process with different number of context service providers and counted the time used in the selection. 
Figure~\ref{fig:perf-algo} shows the results. From Figure~\ref{fig:perf-algo}, we can see that in general the plotted line fits well with the theoretical complexity of the algorithm ($O(n\log n)$). The algorithm is quite efficient. For example, it takes only about 5ms to execute the algorithm with a context service community having 1000 context providers.

%

In the second experiment, we used the same context service community 
but tested the performance of whole selection process which also includes the communication time between the context service community and the context service providers. As illustrated in Figure~\ref{fig:perf-full}, the plotted line is almost linear and the execution time is at the scale of seconds. For example, the whole process used about 2 seconds for 150 context providers, and about 4 seconds for 350 context providers, and 10 seconds for 1000 context providers. We believe that the main reason that why the trajectory is linear is because the communication time increases linearly with the increase of the number of context providers, while the time used on executing the selection algorithm is not significant, which is at the level of milliseconds.

From above experiments, we can draw the conclusion that the performance of current implementation of the context service community manager is acceptable for small to medium size context service communities with hundreds of context service providers.

\subsection{Performance Study of the MoDAR Middleware}
As discussed in Section~\ref{sec:MoDAR}, MoDAR brings the capability of dynamic adaptation to BPEL processes to ContextServ and clearly also brings some performance overhead. We evaluate MoDAR's performance in this section. Based on the anatomy of the MoDAR runtime environment illustrated in Figure~\ref{fig:runtime}, we can see that the performance overhead mainly comes from two places.
Firstly, the aspect service is an extra Web service that needs to be invoked for each aspect\footnote{If an activity has both a before aspect and an around aspect, they will be combined to one aspect Web service invocation at runtime.}, and there are variables exchanged between the process and the aspect service. 
Secondly, every rule is actually the externalization of a conditional statement in the process, and executing a rule may be slower than executing the equivalent conditional statement. 

Several experiments have been conducted to evaluate the actual performance impact of the approach. The MoDAR runtime environment was run on a PC with an Intel Core i7 860 2.80 GHz CPU and 4GB of RAM. Riftsaw 2.2.0\footnote{http://www.jboss.org/riftsaw/} running on top of JBoss-Application-Server-5.1.0GA was used as the BPEL engine, with Drools-5.0\footnote{http://www.jboss.org/drools/} used as the rule engine. 

\begin{figure}[]
	\centering
	\includegraphics[scale=.5]{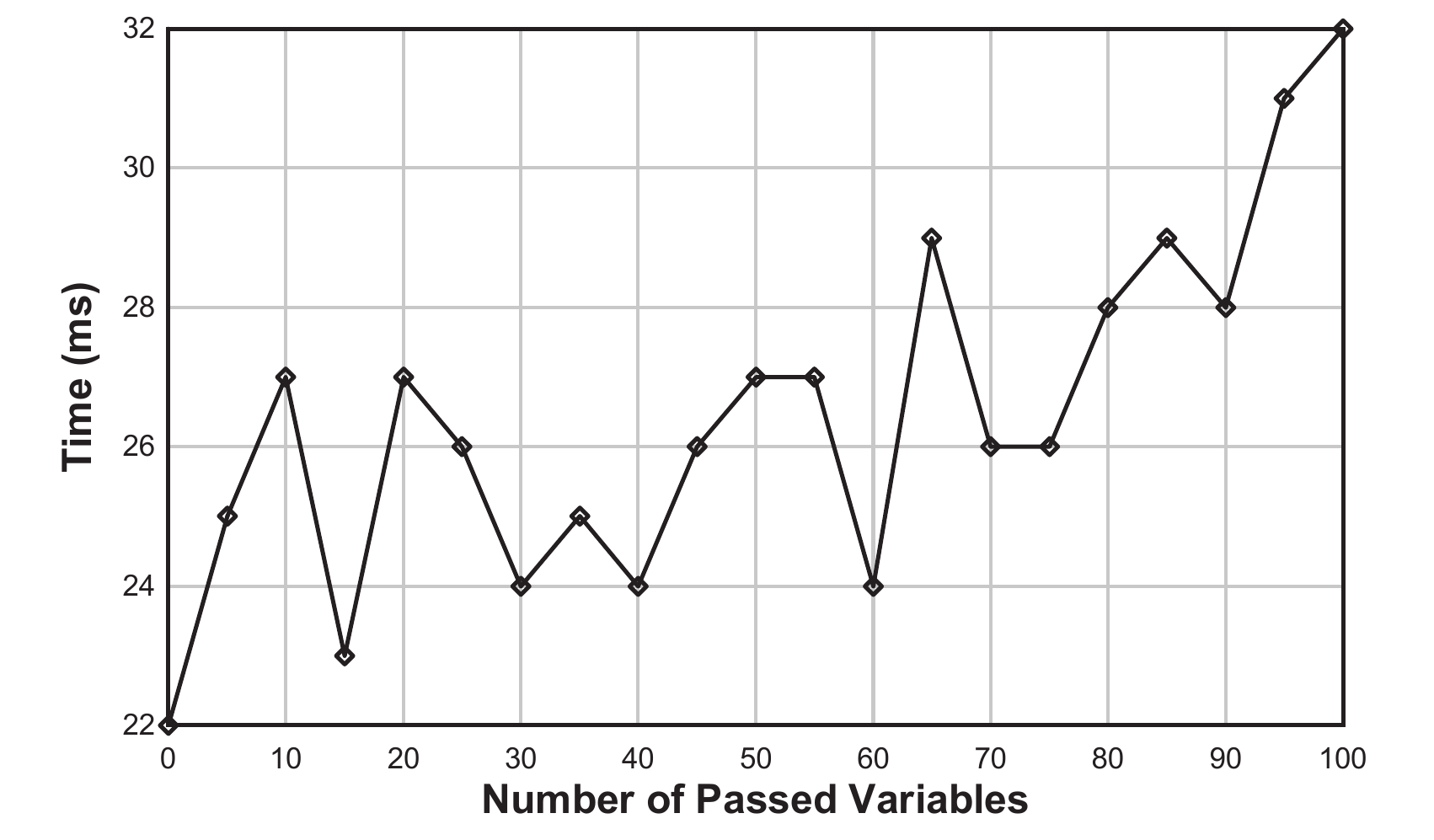}
	\caption{Execution time of a single aspect service w.r.t. the number of passed variables}
	\label{fig:exp1}
\end{figure}

In the first experiment, we tested the impact of invoking a single empty aspect service with various number of randomly generated primitive type variables passed. Every setting was tested five times and the average execution time of an empty aspect service w.r.t. the number of passed variables is shown in Figure~\ref{fig:exp1}. As we can see, it costs 22 ms to invoke an empty aspect service without passing any variables and costs 32 ms to invoke an empty aspect service with 100 variables passed to it. This result shows that the variable exchange between the Riftsaw BPEL server and the Drools server is quite fast, and there is only 10 ms increase from passing no variable to passing 100 variables. The reason could be that these two servers are two components that both run inside the same JBoss application server.

In the second experiment, we compared the performance of an aspect and its equivalent BPEL conditional statements with various number of rules at various conditional expression complexity level. We fixed the number of exchanged variables to 100, with each variable being assigned a random primitive type and value. According to the result of the first experiment, passing 100 variables brings about 10 ms overhead to the overall aspect service invocation time. Because the action part of a rule (e.g., Web service invocation) takes the same time to run no matter it is invoked from a rule engine or from a process engine, we used empty action for all the rules and conditional statements. In addition, we took the number of variables evaluated in a conditional expression as the complexity level. For example, at complexity level 10, 10 variables are randomly selected from the 100 candidates and the value of each variable is tested both in the condition part of a rule and in the corresponding BPEL {\texttt <if>} statement. In the experiment, we selected 1, 10, and 30 as the representative complexity levels, and every setting was tested 5 times.

\begin{figure}[!t]
	\centering
	\subfigure[]{
		\includegraphics[scale=.41]{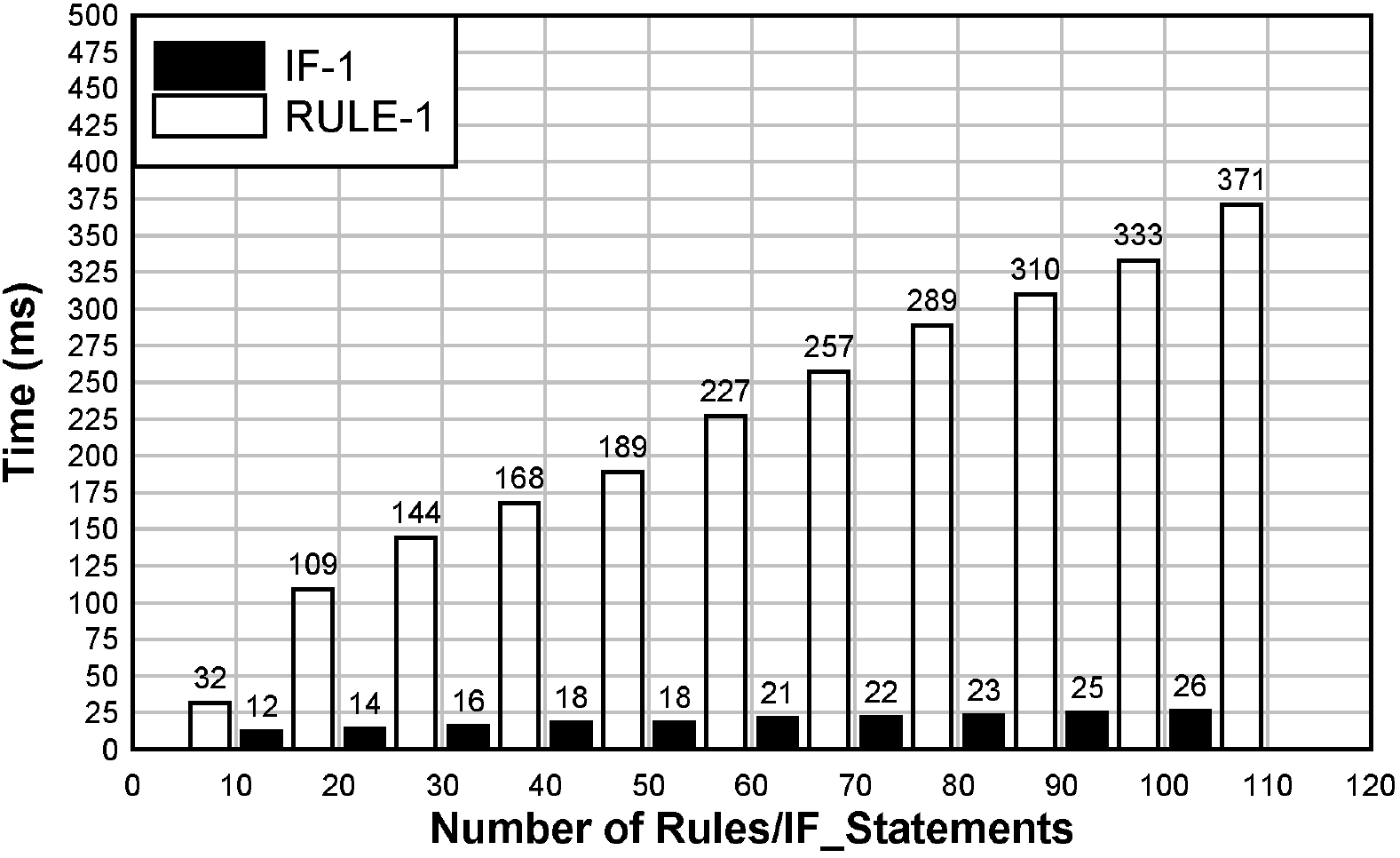}
		\label{fig:subfig1}
	}
	\subfigure[]{
		\includegraphics[scale=.41]{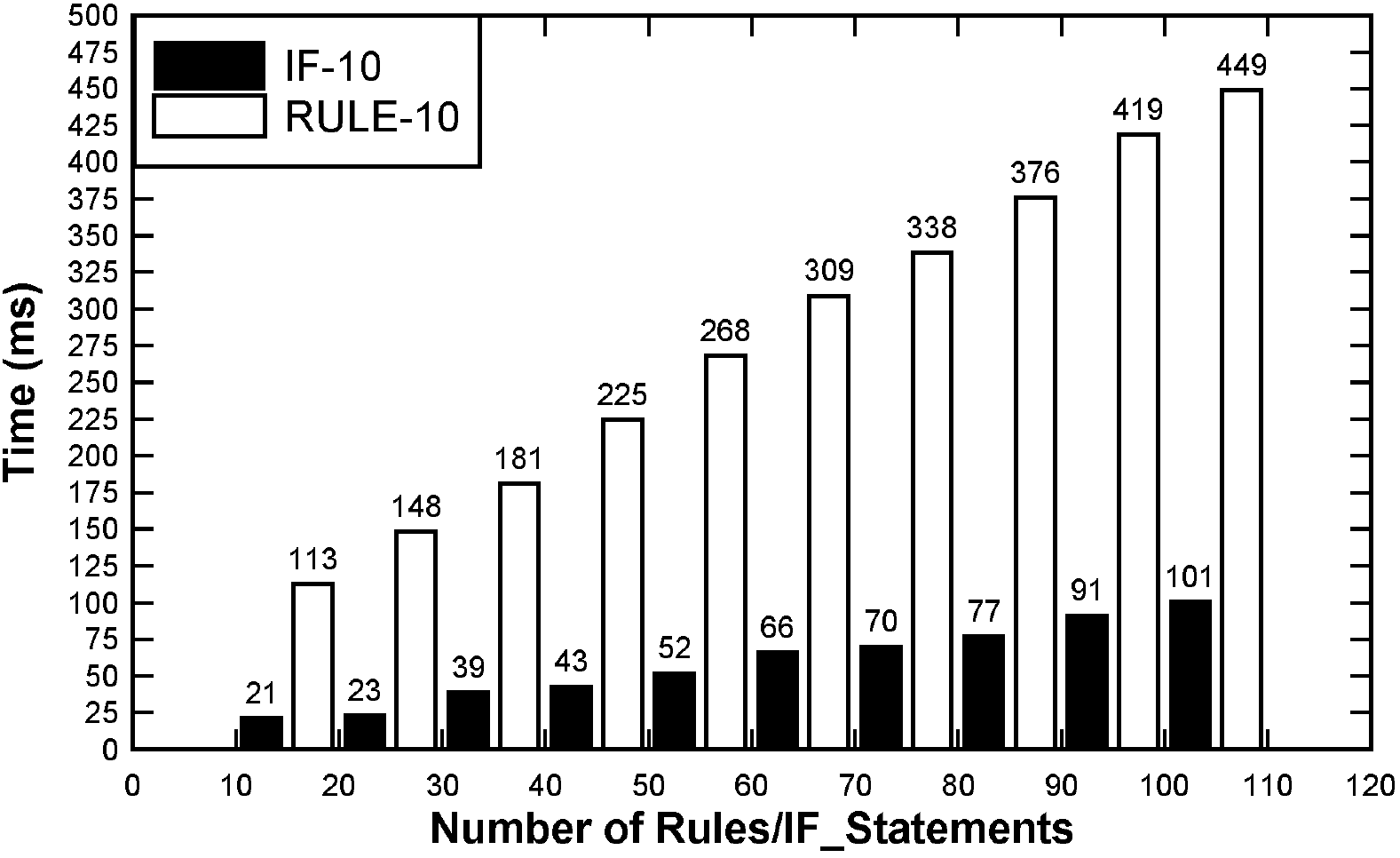}
		\label{fig:subfig2}
	}
	\subfigure[]{
		\includegraphics[scale=.41]{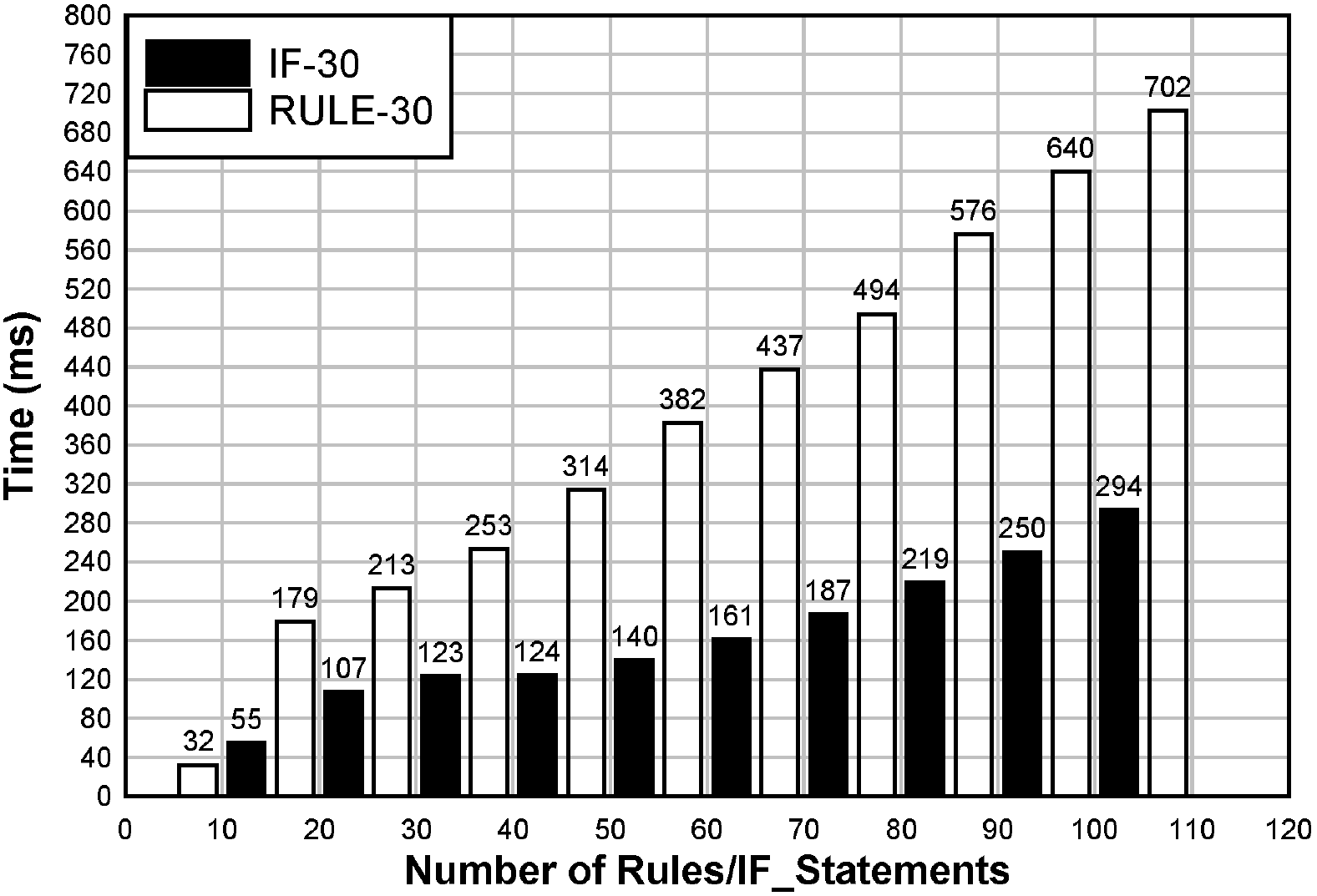}
		\label{fig:subfig3}
	}
	\caption{Performance comparison w.r.t. the number of rules/IF\_statements}
	\label{fig:exp2}
\end{figure}

The result of this experiment is shown in Figure~\ref{fig:exp2}. As we can see, replacing an aspect with its equivalent IF statements always brings some performance overhead. At complexity level 1, the average overhead is 97 ms for 10 rules (i.e., 109 ms for invoking the aspect service minus 12 ms for executing the IF statements), and 345 ms for 100 rules. 
At complexity level 30, the overhead is 124 ms for 10 rules and 408 ms for 100 rules. From this experiment we can see that the overhead of invoking an aspect service varies from 97 ms for a simple case to about 408 ms for a very complex case. 


\section{Discussions and Related Work}
\label{sec:relatedwork}

With the maturing and wide-adopting of Web service technology, research on providing engineering approaches to facilitate the development of context-aware services has gained significant momentum. Using model-driven paradigm to develop CAS has been proven to be a valuable and important strand in this research area considering the quality and efficiency it brings along. Apart from model-driven approaches, in the survey on context-aware service engineering~\cite{Kapitsakietal.2009}, the authors propose other 5 categories of approach: \textit{Middleware solutions and dedicated service platforms}, \textit{Use of ontologies}, \textit{Rule-based reasoning}, \textit{Source code level programming/Language extensions}, and \textit{Message interception}. In general, we agree with their viewpoint that any of the approaches has its pros and cons. For example, the source code level approach can give more freedom to developers to do all kinds of context-aware adaptation, but this approach does not separate apart the concerns on context-awareness and suffers from a significant maintenance cost. As to the model-driven approach, apart from its advantages, it requires to keep the consistency between high level models and low level executable code at all times, which brings extra complexity. We also agree that some approaches can be used at the same time to bring extra benefits. For example, we have adopted ontologies in the context service community to provide enhanced context organization and matching functionality.

In the literature on modeling languages and model-driven development of context-aware services, the following research work relates to ContextServ in particular.

In~\cite{Ayed2006}, Ayed et al. proposed a UML metamodel that supports context-aware adaptation of service design from structural, architectural and behavioral perspectives. The structural adaptation can extend the service object’s structure by adding or deleting its methods and attributes. The architectural adaptation can add and delete service objects of an application according to the context. The behavioral adaptation can adapt the behavior of the service object by extending its UML sequence diagram with optional context related sequences. Furthermore, based on the UML metamodel, in ~\cite{Ayed2007MDD}, Ayed ed al. proposed an MDD approach to model context-aware applications independently from the platform, which includes six phases that approach step by step the mechanisms required to acquire context information and perform adaptations.

In~\cite{Grassi2007,Sindico2009}, Sindico and Grassi proposed CAMEL (Context Awareness ModEling Language) which considers both model-driven development and aspect-oriented design paradigms so that the design of the application core can be decoupled from the design of the adaptation logic. In particular, CAMEL categorizes context into \texttt{state-based} which characterizes the current situation of an entity and \texttt{event-based} which represents changes in an entity’s state. Accordingly, state constraints, which are defined by logical predicates on the value of the attributes of a state-based context, and event constraints, which are defined as patterns of event, are used to specify context-aware adaptation feature of the application. 

In~\cite{Malek2010}, Malek et al. proposed CAAML (Context-aware Adaptive Activities Modeling Language) which aims at modeling context-aware adaptive learning activities in the E-learning domain. This language focuses on modeling two classes of rules: rules for context adaptation and rules for activity adaptation to support pedagogical designing.

In~\cite{hoyos2010mlMLcontext,hoyos2013domain} Hoyos et al. proposed a textual Domain-Specific Language(DSL), namely MLContext, which is specially tailored for modeling context information. It has been implemented by applying MDD techniques to automatically generate software artifacts from context models. The MLContext abstract syntax has been defined as a metamodel, and model-to-text transformations have been written to generate the desired software artifacts(e.g. OCP middleware and JCAF middleware). The concrete syntax has been defined with the EMFText tool, which generates an editor and model injector. Furthermore, in~\cite{hoyos2016model}, MLContext is extended for modeling QoC and the models can be mapped to code for two frameworks(COSMOS and SAMURA) supporting QoC.

In~\cite{Prezerakos2007}, the authors' work addresses the decoupling of core service logic from context-related functionality by adopting a model-driven approach based on a modified version of ContextUML. Core service logic and context handling are treated as separate concerns at the model level as well as in the resulting source code. In design phase, besides class diagrams, UML activity diagrams are used for modeling the core service logic flow in conjunction with MDE (Model-driven Engineering) transformation techniques and AOP (Aspect Oriented Programming). In coding phase, AOP encapsulates context-dependent behaviors in discrete AspectJ code modules. Context binding information provided in UML models is used to create pointcuts and related advices, as well as to create the binding between them. 

In~\cite{kapitsaki2009model}, Kapitsaki et al. proposed an architecture for the context adaptation of web applications consisting of web services and a model-driven methodology for the development of such context-aware composite applications. In the methodology, the web application functionality is completely separated from the context adaptation at all development phases (analysis, design and implementation). In the modeling level, composite web applications are modeled in UML and the application design is kept, at a great extent, independent from specific platform implementations and flexible enough to allow the introduction of different code specific mappings. Context adaptation is performed on a service interface level to keep client independent. The modeling exploits a number of pre-defined profiles, whereas the target implementation is based on an architecture that performs context adaptation of web services based on interception of Simple Object Access Protocol(SOAP) messages. 

In~\cite{boudaa2016aspect}, Boudaa et.al. proposed an approach taking advantage of combining MDD and AOP to sustain the development of context-aware service-based applications in mobile and ubiquitous environments. Contexts are modeled with a proposed ontology-based context model which is structured on three sub-ontologies: generic, domain and application ontologies. A UML-based metamodel, called ContextAspect, is proposed to define and specify where and how the context-aware adaptation takes place. The ContextAspect metamodel is composed of three parts: aspect modeling, context modeling and context-awareness modeling. AOM handles the context-awareness logic in ContextAspect models (as variants) to fill context-aware application elements (as variation points) by using weaving techniques at design and run times. At design-time, the weaving enables to produce a wide range of context-aware application models without designing them from the beginning. The run-time weaving consists of weaving necessary reconfiguration into the running application according to the context change, so accomplishing its dynamic adaptation. 

Table~\ref{tab:MDDcomparison} gives a detailed summary of the above mentioned related works and comparison between them with our approach from the perspectives of modeling language, MDD techniques, tools and platform.
\begin{sidewaystable}[htbp]
	
	\centering 
	\caption{Summary and comparison of model-driven approaches for context-aware application development}
		\begin{scriptsize}

		\begin{tabular}{|l|l|c|c|c|c|c|c|c|c|} 
			\hline
			
			\multicolumn{2}{|c|}{}& \makecell{Ayed \\ 2006,2007} &\makecell{Sindico-Grassi \\2007,2009}& \makecell{Prezerakos \\2007} & \makecell{Kapitsaki \\2009} &\makecell{Malek\\2010}&\makecell{Hoyos\\2013,2016}&\makecell{Boudaa\\2016}& ContextServ \\
			\hline
			
			\multirowcell{5}{Context\\modeling}&\makecell{Modeling language\\(based-Model)}&UML&\makecell{UML\\(CAMEL)}&\makecell{UML\\(ContextUML*)}& \makecell{UML\\(ContextUML*)} & \makecell{UML\\(CAAML)} & \makecell{DSL\\(MLContext)}&\makecell{Ontology\&\\UML}& \makecell{UML\\(ContextUML)} \\ \cline{2-10}
			&Atomic context & + & + & + & + & + & + & +  & + \\ \cline{2-10}
			&Composite context & - & + & + & + & + & + & +  & + \\ \cline{2-10}
			&Context quality& + & - & - & - & - & + & - & + \\ \cline{2-10}
			&Context sensing &+ & - & - & - & - & + & + & -\\ 
			\hline
			
			\multicolumn{2}{|c|}{Service modeling}&- &- & + & + & - & - & - & +\\ 
			\hline

			\multirowcell{3}{Context-awareness\\modeling}&Context binding&+&+&+&+&+&+&+&+ \\ \cline{2-10}
			&Context triggering&+&+&+&+&+&+&+&+ \\\cline{2-10}
			&Behavior adaptation&+&-&-&-&-&-&+&+ \\
			\hline
			
			\multicolumn{2}{|c|}{Decoupling business logic and context logic}&+ &+ & + & + & + & + & + & +\\ 
			\hline
			
			\multicolumn{2}{|c|}{Adaptation time (design-time/run-time)}&design-time &\makecell{design-time\\run-time} & \makecell{design-time\\run-time} & \makecell{design-time\\run-time} & design-time & design-time & \makecell{design-time\\run-time} & \makecell{design-time\\run-time}\\ 
			\hline
			
			\multicolumn{2}{|c|}{Implementation platform}&unspecified &AspectJ & SOA & SOA & IMS-LD & OCP/JCAF & \makecell{SOA\\(FraSCAti)} & \makecell{SOA\\(BPEL)}\\ 
			\hline
			
			\multirowcell{2}{Supporting software\\tools}&\makecell{Graphical modeling\\ environment}&- &+ &+ &- &- &+ &+ &+ \\ \cline{2-10}
			&Transformation tool&- &- &+ &+ &- &+ &+ &+ \\ 
			\hline

		\end{tabular}	
			
		\end{scriptsize}
		
	
	\label{tab:MDDcomparison}
\end{sidewaystable}

As we have discussed in detail in Section~\ref{sec:ContextUML}, ContextUML is a UML language for model-driven development of CASs. The significant features of ContextUML include:

\begin{itemize}
	\item {\em Context modelling}: supports both atomic contexts modelling and state-chart-based composition of atomic contexts; context quality can be specified and quality-based context selection is supported by the Context Community middleware.
	\item {\em Service modelling}: has dedicated UML meta-classes to model the structure of Web services.
	\item {\em Context-awareness modelling}: provides two mechanisms-context binding for automatic context retrieval, and context triggering for automatic context-aware adaptation.
\end{itemize}

To the perspective of modelling language for context-aware application development, we compare ContextUML with the other metamodels from the issues of context modelling, servcie modelling, and context-awareness modelling. It should be noted that, in~\cite{Prezerakos2007,kapitsaki2009model}, their models are modified versions of ContextUML, so most of the language capabilities of their models equal to ContextUML's and the comparison with them will not discussed below. As we can see from the table, all languages support the modelling of atomic context. For composite context, although CAMEL and CAAML claim that atomic contexts can be aggregated but no details were given in the papers. In~\cite{boudaa2016aspect}, composite context is inferred from low-level contexts using Semantic Web Rule Language (SWRL), and in MLContext, simple references are used to link composite context with their atomic contexts. ContextUML gives a complete approach to composing a composite context from atomic contexts in statechart which is a widely used formalism integrated into UML. ContextUML supports context quality modelling and use context service community to support QoC-based context selection. Although Ayed UML is able to specify the quality attributes of a context, no runtime support was reported in the paper. In~\cite{hoyos2016model}, MLContext was extended for modelling QoC, however, it doesn't support QoC-based context selection. In~\cite{Prezerakos2007,kapitsaki2009model}, ContextUML is used in modified versions without context service community, so context quality modelling is not supported. ContextUML does not directly support the modelling of context sensors but the information of a context sensor is abstracted in a ContextService element and UML models on sensors can be seamlessly integrated using the ConttextService element.

As to service modelling, only ContextUML directly supports the structure of Web services, which is of enormous importance to the development of context-aware Web services. The other languages just use plain UML classes to represent Web services or even without support of Web services. 

For context-awareness modelling, all the languages except MLContext support the main features including context binding and context triggering. However, only ContextUML (integrated  with MoDAR), Ayed UML and ContextAspect model support behavior adaptation, which means a service or process has the ability to change its behavior at runtime in accordance with the changes in the requirements and/or the external environment(contexts). In our MoDAR models, the base model captures the flow logic of the requirements, the variable model captures the decision logic of the requirements, and the weave model weaves the base model and variable models together using an aspect-oriented mechanism so that runtime changes can be applied to the variable model without affecting the base model. Dynamic behavior adaptation is achieved in a way that we can freely add/remove/replace business rules defined in the modelling phase and then transform and redeploy them without terminating the execution of the process. Ayed UML only supports to define behavior adaptation in design-time, and no run-time support is reported in the paper. Both of MoDAR and ContextAspect model support behavior adaptation in design-time and run-time. The mechanism of behavior adaptation in ContextAspect model enables to change alternatively the application behavior by selecting one among several behaviours in accordance with current contextual situation. 
By contrast, MoDAR's ontology-based rule language in variable model supports rule types of $constraint$, $computation$, $inference$, $action$, which can provide more powerful adaptivity and is applicable to more contextual situations. 

Because dynamic adaptation is closely related to the targeting system, we also listed the supported targeting implementation platform of each approach. Different implementation languages or underlying frameworks/platforms and middleware are adopted in each approach. ContextServ and approaches presented in~\cite{Prezerakos2007,kapitsaki2009model,boudaa2016aspect} support the SOA paradigm. As discussed in Section~\ref{sec:Transformer} and~\ref{sec:MoDAR platform}, after modelling adaptation in ContextUML and MoDAR's models, ContextServ can transform the Web services and the behavior adaptation will be reflected in standard BPEL that has become a de facto industry standard (widely adopted by major IT service providers including IBM, Oracle, and SAP) to create composite service processes and applications. ~\cite{boudaa2016aspect} uses FraSCAti platform as the target platform which supports Service Component Architecture(SCA). Models of MLContext can be transformed to specific context middleware(e.g., OCP and JCAF). Because CAAML is a modelling language in the E-learning domain, IMS-LD (IMS Learning Design)\footnote{http://www.imsglobal.org/learningdesign/index.html} is the targeting system but the detail of transformation is not report in the paper. CAMEL is still an ongoing work, so only examples on how to transform to ContextJ~\cite{Hirschfeld2008} were described in the paper. As to Ayed UML, no targeting systems are reported in the paper.

For supporting software tools, ContextUML has a comprehensive graphical modelling environment developed on top of ArgoUML and also a full-fledged automatic transformation tool for generating deployable BPEL code. All of CAMEL, CAAML, MLContext and language in~\cite{boudaa2016aspect} have a graphical modelling environment based on Eclipse EMF\footnote{http://www.eclipse.org/modelling/emf/}, but no fully workable transformation tools are reported of CAMEL and CAAML.

To summarize, ContextUML has the richest set of features to support the development of CASs among all the discussed modelling languages. And ContextServ provides a comprehensive platform where context-aware Web services are specified in a high-level modelling language and their executable implementations are automatically generated and deployed, thus contributing significantly to both design flexibility and cost savings. 

\section{Conclusions}
\label{sec:conclusion}

In recent years, CASs are emerging as an important technology for building innovative context-aware applications. Unfortunately, CASs are still difficult to build, due to lack of context provisioning management approach and lack of generic approach for formalizing the development process.

In this paper, we have introduced ContextUML, a UML-based modelling language, and the ContextServ platform that implements ContextUML, for model-driven development of CASs. We have also introduced the context service community middleware and the MoDAR middleware. 
The former facilitates dynamic and optimized context provisioning for CASs, and the latter supports dynamic adaption of CASs at runtime. The evaluation of ContextServ as a development tool and as a runtime environment, and the comparison with related work show that ContextServ can support effective development and efficient execution of context-aware Web services. 
  
Our future work will focus on evaluating the applicability of ContextServ in the flourishing IoT domain considering 
that 
 context-aware computing plays an important role to the the success of this emerging technology~\cite{PereraLJC15}, and propose new context modelling, management, and context-aware adaptation techniques that cater for the characteristics of IoT such as large-scale-ness and dynamicity.

\begin{acks}
The authors would like to thank Sam Pohlenz, Hoi Sim Wong, Kewen Liao for their participation in the implementation of ContextServ.
\end{acks}

\bibliographystyle{ACM-Reference-Format}
\bibliography{contextserv}

\end{document}